\documentclass[nofootinbib]{article}

\usepackage{amssymb}

\usepackage{amsmath}

\usepackage{bm}

\usepackage{jcappub}

\usepackage{xcolor}

\allowdisplaybreaks

\newcommand{\beq}{\begin{equation}}
\newcommand{\eeq}{\end{equation}}
\newcommand{\bea}{\begin{eqnarray}}
\newcommand{\eea}{\end{eqnarray}}
\newcommand{\ben}{\begin{enumerate}}
\newcommand{\een}{\end{enumerate}}

\newcommand{\pa}{\partial}

\newcommand{\ed}{{\rm d}}
\newcommand{\ced}{{\rm D}}
\newcommand{\we}{\wedge}

\newcommand{\Ord}{{\cal O}}

\newcommand{\Tr}{{\rm Tr}}
\newcommand{\ti}{\tilde}

\newcommand{\what}{\widehat}

\newcommand{\Rs}{\mathbb{R}}
\newcommand{\Cs}{\mathbb{C}}

\renewcommand\({\left(}
\renewcommand\){\right)}
\renewcommand\[{\left[}
\renewcommand\]{\right]}
\newcommand{\bra}{\langle}
\newcommand{\ket}{\rangle}
\newcommand{\Bra}{\left\langle}
\newcommand{\Ket}{\right\rangle}

\newcommand{\nn}{\nonumber}

\newcommand{\al}{\alpha}
\newcommand{\be}{\beta}
\newcommand{\ga}{\gamma}
\newcommand{\Ga}{\Gamma}
\newcommand{\de}{\delta}

\newcommand{\vep}{\varepsilon}

\newcommand{\et}{\eta}

\newcommand{\ka}{\kappa}
\newcommand{\la}{\lambda}
\newcommand{\La}{\Lambda}
\newcommand{\ro}{\rho}
\newcommand{\si}{\sigma}

\newcommand{\ph}{\phi}

\newcommand{\ch}{\chi}

\newcommand{\cA}{{\cal A}}

\newcommand{\cS}{{\cal S}}

\newcommand{\cY}{{\cal Y}}
\newcommand{\cX}{{\cal X}}

\newcommand{\cC}{{\cal C}}

\newcommand{\cF}{{\cal F}}
\newcommand{\cW}{{\cal W}}

\title{Spin connection formulations of real Lorentzian General Relativity}

\author{Ermis Mitsou} 
\emailAdd{ermitsou@physik.uzh.ch}
\affiliation{Center for Theoretical Astrophysics and Cosmology, Institute for Computational Science, University of Zurich, CH--8057 Z\"urich, Switzerland}

\abstract{We derive the pure spin connection and constraint-free BF formulations of real four-dimensional Lorentzian vacuum General Relativity. In contrast to the existing complex formulations, an important advantage is that they do not require the reality constraints that complicate quantization. We also consider the corresponding modified gravity theories and point out that, contrary to their self-dual analogues, they are not viable because they necessarily contain ghosts. In particular, this constrains the set of potentially viable unified theories one can build by extending the gauge group to the ones with the action structure of General Relativity. We find, however, that the resulting theories do not admit classical solutions. This issue can be solved by introducing extra dynamical fields which, incidentally, could also provide a way to include a matter sector.}

\begin{document}

\maketitle

\flushbottom

\section{Introduction \& summary}

Alternative formulations of classical General Relativity (GR) provide us with important insights about the theory, new perspectives for potential modifications or extensions and also genuinely new starting points for approaching the quantum theory. One line of research in this direction are the so-called ``pure connection" formulations, where the usual gravitational field that is a metric or a vierbein, is replaced by a connection associated with some group, in close analogy with the mathematical description we have for the rest of the known forces of nature. In fact, this suggestive resemblance is an important motivation for this approach, because enlarging the gauge group leads to extra connection components and therefore constitutes an elegant potential path towards unification (see the review \cite{KP2017} and references therein). If one considers the metric formalism, then the only available group is the diffeomorphism group, so extending it necessarily introduces extra dimensions. In the vierbein formalism, however, we have the internal action of the Lorentz group, so the latter can be extended without altering the dimensionality of space-time.

Nevertheless, it is instructive to quickly expose the metric case, whose development dates back to Eddington \cite{Eddington} and Schr\"odinger \cite{Schrodinger}. The starting point is the Hilbert-Palatini action for a metric $g_{\mu\nu}$ and an independent symmetric affine connection $\Ga^{\ro}_{\mu\nu} \equiv \Ga^{\ro}_{\nu\mu}$ 
\beq
S = \frac{1}{16\pi G} \int \ed^4 x\, \sqrt{-g} \( g^{\mu\nu} R_{\mu\nu} - 2\La \) \, ,
\eeq
where 
\beq
R_{\mu\nu} := \pa_{\ro} \Ga^{\ro}_{\mu\nu} - \pa_{(\mu} \Ga^{\ro}_{\nu)\ro} + \Ga^{\ro}_{\ro\si} \Ga^{\si}_{\mu\nu} - \Ga^{\ro}_{\si\mu} \Ga^{\si}_{\ro\nu} \, .
\eeq
Integrating out $\Ga^{\ro}_{\mu\nu}$, i.e. replacing it with the solution to its equation of motion, yields the Einstein-Hilbert action. On the other hand, integrating out $g_{\mu\nu}$ leads to the pure connection action
\beq \label{eq:SPCES}
S \to \frac{1}{8\pi G \La} \int \ed^4 x\, \sqrt{- \det R_{\mu\nu}}   \, .
\eeq
Note that this manipulation requires $\La \neq 0$, which fits within the present concordance viewpoint on cosmology and leads in particular to a tiny dimensionless ``coupling constant" $\sqrt{8\pi G \La} \sim 10^{-60}$ in $\hbar = c = 1$ units. 

In the vierbein formalism, the situation is more involved. If one starts with the analogous Hilbert-Palatini formulation, with the independent connection being now associated with the Lorentz group (the ``spin" connection), then it is easy to see that the vierbein cannot be isolated in its equation of motion in an algebraically simple manner. A more fruitful starting point is rather the Plebanski action \cite{Plebanski}, which has led to a formulation involving the spin connection and an auxiliary scalar density by Capovilla, Jacobson and Dell \cite{CJD1989, CJD1991, CJD1992} (see also \cite{Peldan, CJ1992}) and more recently to a formulation involving the spin connection alone for the $\La \neq 0$ case by Krasnov \cite{Krasnov2011a, Krasnov2011b} (see also \cite{DKS2012, DKS2012b, GKS2013, DKS2014} for the corresponding quantum field theory). One common feature of these formulations is that they depend on the self-dual component of the spin connection alone. They can therefore be seen as gauge theories of the complexified group SO$(3,\Cs)$, commonly referred to as ``complex GR", and thus require ``reality" constraints on the curvature field in order to select the sector corresponding to real Lorentzian GR. These constraints depend on the action and field content and are especially difficult to handle in the quantum theory. We will refer to these formulations as the ``self-dual" ones (SD). 

In the case of real GR, where the full spin connection is considered, the algebraic structure is more complicated and this fact has prevented the derivation of the pure connection action so far. In this paper we will show how to overcome this difficulty and will therefore obtain the pure spin connection formulation of vacuum GR with $\La \neq 0$. As in the SD case \cite{Krasnov2011a, Krasnov2011b}, the action involves a matrix square root
\beq \label{eq:SPC}
S_{\rm GR}[F] := \frac{1}{16\pi G \La} \int \Tr \[ \sqrt{\bm{z} \bm{\cX} \bm{z}} \]^2 \, ,
\eeq
where
\beq \label{eq:Omega}
\cX^{ab}_{\,\,\,\,\,\,cd} := F^{ab} \we F_{cd} \, , \hspace{1cm} \bm{z}^{ab}_{\,\,\,\,\,\,cd} := \frac{\( \ga + 1 \) \( \de^a_c \de^b_d - \de^a_d \de^b_c \) + \( \ga - 1 \) \vep^{ab}_{\,\,\,\,\,\,cd}}{2\ga} \, ,  
\eeq
are seen as a $6 \times 6$ matrices in antisymmetric pairs of Lorentz indices $[ab]$, $F^{ab}$ are the curvature 2-forms, $\ga$ is the Immirzi parameter, $\vep_{abcd}$ is the Levi-Civita symbol and the indices are displaced with the Minkowski metric $\et_{ab}$. Note that in the $\La = 0$ case \eqref{eq:SPC} can be expressed non-singularly with the help of an extra Lagrange multiplier density, but the corresponding dynamics are not well defined, contrary to the SD formulation \cite{CJD1991}. 

Another useful approach to gauge theories is through the so-called ``BF" or ``covariant Hamiltonian" formulation, where one integrates in an auxiliary set of 2-forms $B^{ab}$ to make the equations of motion first-order in derivatives. The best known BF formulation of GR is the Plebanski action, in both complex \cite{Plebanski} and real \cite{DPF1998} forms, in which case one requires an extra set of Lagrange multipliers in order to impose some conditions on $B^{ab}$ known as the ``simplicity" or ``metricity" constraints. Recently, however, it was noted by Herfray and Krasnov \cite{HK2015,Krasnov2017} that there exists a constraint-free BF formulation of GR, at least in the real case, since the complex one still requires the reality constraints. This description is therefore closer to the structure of standard gauge theory, which is again an important feature if one is interested in unified theories. Due to the same kind of algebraic complications that prevented the derivation of the real pure connection action, the real version of this BF formulation has not yet been given either. Here we derive this action too, with the result being
\beq \label{eq:SBF}
S_{\rm GR}[B,F] := \frac{1}{2} \int \[ B_{ab} \we F^{ab} + \Tr \( \bm{w} \bm{\cY} \bm{w} \) + \frac{\[ \Tr \sqrt{\bm{z} \bm{w}^2 \bm{\cY} \bm{w}^2 \bm{z}} \]^2}{8\pi G \La - \Tr \[ \( \bm{z} \bm{w} \)^2 \]} \] \, ,
\eeq
where now 
\beq
\cY^{ab}_{\,\,\,\,\,\,cd} := B^{ab} \we B_{cd} \, ,  \hspace{1cm} w^{ab}_{\,\,\,\,\,\,cd} := w \( \de^a_c \de^b_d - \de^a_d \de^b_c \) + \ti{w}\, \vep^{ab}_{\,\,\,\,\,\,cd} \, ,
\eeq
and $w$, $\ti{w}$ are two additional free parameters. In contrast to the pure connection formulation \eqref{eq:SPC}, here we do have access to the $\La = 0$ case. 

By expressing \eqref{eq:SPC} and \eqref{eq:SBF} in their SL$(2,\Cs)$ form, one can recover the SD formulations \cite{Krasnov2011a, Krasnov2011b,HK2015,Krasnov2017} either by simply setting $\ga = -i$, or by imposing the appropriate reality constraints. In doing so, we will clarify how an implicit sign ambiguity in the scalar reality constraint, for the pure connection case, is related to the two sectors of the underlying Plebanski theory. Another interesting computation in this paper is the linearization of the pure connection theory around (anti-)de-Sitter space-time ((A)dS). We perform it for generic $\ga$ values and find agreement with the earlier perturbative constructions for the non-chiral case $|\ga| \to \infty$ \cite{Zinoviev, BBB2015}. A special aspect of the quadratic action is that it only depends on the fully traceless part of the curvature tensor and we show that this is a peculiarity of the GR action. 

Finally, we will also consider the known generalizations of \eqref{eq:SPC} and \eqref{eq:SBF} where the matrix function $f(\bm{M}) = [ \Tr \sqrt{\bm{M}} ]^2$ is replaced by an arbitrary homogeneous function of degree one \cite{Bengtsson1990,Krasnov2006,Bengtsson2007,Krasnov2007,Krasnov2008b,Krasnov2008,Freidel2008,Krasnov2009,Krasnov2012}. We will argue that, contrary to the SD case, these modified gravity theories have an unavoidable pathology in that one of their two gravitons is necessarily a ghost. This fact, which does not seem to be noted in the literature to our knowledge, makes it all the more important to know the precise matrix functions \eqref{eq:SPC} and \eqref{eq:SBF} that correspond to GR. This is also relevant for the corresponding unified theories, i.e. those obtained by extending the gauge group ${\rm SO}(1,3) \to G$, because then only the GR Lagrangian function has a chance of yielding viable results. We will show, however, that the corresponding equations of motion admit no solution as soon as the dimension of $G$ exceeds six. This situation could be avoided by introducing extra dynamical fields which, incidentally, could also allow one to describe the matter sector.

The paper is organized as follows. In section \ref{sec:notation} we define a compact matrix notation along with an algebraic toolkit that will make our computations much more straightforward and transparent. In section \ref{sec:Plebanski} we point out the equivalence between a nine-parameter family of real Plebanski formulations, which is useful because some of them can be more suited than others, depending on the problem at hand. We then derive the pure-connection action in section \ref{sec:PC} and the constraint-free BF action in section \ref{sec:BF}. In section \ref{sec:chiral} we show the relation with the SD formulations and in section \ref{sec:linear} we derive the linearized pure connection theory. Finally, in section \ref{sec:ghost} we discuss the ghost problem of the modified gravity theories and in section \ref{eq:unif} we consider some aspects of the group extension of the GR action.

\section{Notation and useful identities} \label{sec:notation}

We will work with tensors in the Lorentz algebra that are therefore indexed by antisymmetric pairs of Lorentz indices $[ab]$ forming a six-dimensional index. More precisely, there will be vectors $V^{[ab]}$ and 2-tensors $M^{[ab][cd]}$ and the metric displacing these pairs of indices is the Killing form of the algebra
\beq \label{eq:Killing}
\ka_{[ab][cd]} := \et_{ac} \et_{bd} - \et_{ad} \et_{bc} \, .
\eeq
We can therefore introduce a compact notation in which $V^{ab} \to | V \ket$ is a vector and $M^{ab}_{\,\,\,\,\,\,cd} \to \bm{M}$ is a matrix. The product and trace definitions must then include $1/2$ factors for each contraction
\beq
(\bm{M} | V \ket)^{ab} := \frac{1}{2}\, M^{ab}_{\,\,\,\,\,\,cd} V^{cd} \, , \hspace{1cm} (\bm{M}^2)^{ab}_{\,\,\,\,\,\,cd} := \frac{1}{2}\, M^{ab}_{\,\,\,\,\,\,ef} M^{ef}_{\,\,\,\,\,\,cd} \hspace{1cm} \bra \bm{M} \ket := \frac{1}{2}\, M^{ab}_{\,\,\,\,\,\,ab} \, ,
\eeq
to avoid counting every independent component twice. The Lorentz-compatible transposition operation is 
\beq
| V \ket^{\ka} := | V \ket^T \ka \equiv \bra V | \, ,  \hspace{1cm} \bra V |^{\ka} \equiv | V \ket \, , \hspace{1cm} \bm{M}^{\ka} := \ka^{-1} \bm{M}^T \ka \, ,
\eeq
which therefore yields covectors $V_{ab}$ when acting on vectors $V^{ab}$, and vice-versa, and exchanges the pairs of indices for tensors $(\bm{M}^{\ka})_{abcd} \equiv M_{cdab}$. We will refer to it simply as ``transposition" and note that it has all the algebraic properties of the usual matrix transposition. In particular, we can define the corresponding ``symmetric" and ``antisymmetric" matrices as the ones satisfying
\beq
\bm{S}^{\ka} \equiv \bm{S} \, , \hspace{1cm} \bm{A}^{\ka} \equiv - \bm{A} \, ,
\eeq
respectively, and any matrix $\bm{M}$ can be decomposed as
\beq
\bm{M} \equiv \bm{S} + \bm{A} \, , \hspace{1cm} \bm{S} := \frac{1}{2} \( \bm{M} + \bm{M}^{\ka} \) \, , \hspace{1cm} \bm{A} := \frac{1}{2} \( \bm{M} - \bm{M}^{\ka} \) \, .
\eeq
We can also define two Lorentz-invariant inner products and an outer one
\beq \label{eq:inoutprods}
\bra V | W \ket \equiv \bra W | V \ket \equiv \frac{1}{2}\, V_{ab} W^{ab} \, , \hspace{0.8cm} \bra \bm{M}^{\ka} \bm{N} \ket \equiv \frac{1}{4} \, M_{abcd} N^{abcd} \, , \hspace{0.8cm} \( | V \ket \bra W | \)^{ab}_{\,\,\,\,\,\,cd} \equiv V^{ab} W_{cd} \, ,
\eeq
with the inner ones being invariant under transposition, while $(| V \ket \bra W |)^{\ka} \equiv |W \ket \bra V|$. There are two Lorentz-invariant matrices, the identity and the dual operator (or ``complex structure")
\beq
\bm{1}^{ab}_{\,\,\,\,\,\,cd} := \ka^{ab}_{\,\,\,\,\,\,cd} \equiv \de^a_c \de^b_d - \de^a_d \de^b_c \, , \hspace{1cm} \star^{ab}_{\,\,\,\,\,\,cd} := \vep^{ab}_{\,\,\,\,\,\,cd} \, ,
\eeq
respectively, which are symmetric and satisfy
\beq
\bm{1} \bm{M} \equiv \bm{M} \bm{1} \equiv \bm{M} \, , \hspace{1cm} \star^2 \equiv - \bm{1} \, , \hspace{1cm}  \bra \bm{1} \ket \equiv 6 \, , \hspace{1cm} \bra \star \ket \equiv 0 \, .
\eeq
With the identity operator one can define the inverse $\bm{M}^{-1}$ of a matrix $\bm{M}$
\beq
\bm{M}^{-1} \bm{M} \equiv \bm{1}  \hspace{1cm} \Leftrightarrow \hspace{1cm} \frac{1}{2}\, (\bm{M}^{-1})^{ab}_{\,\,\,\,\,\,ef} M^{ef}_{\,\,\,\,\,\,cd} \equiv \de^a_c \de^b_d - \de^a_d \de^b_c \, .
\eeq 
Observe that the real matrices of the form $\bm{z} := \al \bm{1} + \be\, \star$ have the algebraic behavior of complex numbers among themselves, e.g.
\beq
\bm{z} \bm{z}' \equiv \bm{z}' \bm{z} \equiv \( \al\al' - \be \be' \) \bm{1} + \( \al \be' + \al' \be \) \star \, , \hspace{1cm} \bm{z}^{-1} \equiv \frac{\al \bm{1} - \be \, \star}{\al^2 + \be^2}  \, ,
\eeq
with the only important difference being that they do not commute with other matrices in general. We will refer to such matrices as the ``invariant" ones. Every symmetric matrix $\bm{S}$ can be decomposed into an invariant part and a covariant one
\beq
\bm{S} \equiv \frac{1}{6} \( S \, \bm{1} - \ti{S} \, \star \) + \what{\bm{S}} \, , \hspace{1cm} S := \bra \bm{S} \ket \, , \hspace{0.5cm} \ti{S} := \bra \star \bm{S} \ket \, , \hspace{1cm} \bra \what{\bm{S}} \ket \equiv \bra \star \what{\bm{S}} \ket \equiv 0  \, ,
\eeq
or, equivalently, into scalar and pseudo-scalar traces and a traceless part. We next note that the presence of a complex structure $\star$ provides a ``conjugation" operation on matrices
\beq
\bm{M}^{\star} :=  - \star \bm{M} \,\star \, ,
\eeq
which is also a symmetry of the matrix inner product \eqref{eq:inoutprods} and of the invariant matrices $\bm{1}$ and $\star$. Moreover, it commutes with transposition $\bm{M}^{\ka\star} \equiv \bm{M}^{\star\ka}$, it is an involution $\bm{M}^{\star\star} \equiv \bm{M}$ and also a similarity transformation $\bm{M}^{\star} \equiv \star^{-1} \bm{M} \, \star$, meaning that it commutes with any analytic matrix function $f(\bm{M})^{\star} \equiv f(\bm{M}^{\star})$. We can then decompose any matrix into its ``even" and ``odd" components
\beq
\bm{M} \equiv \bm{M}_+ + \bm{M}_- \, , \hspace{1cm} \bm{M}_{\pm}^{\star} \equiv \pm \bm{M}_{\pm} \hspace{1cm} \bm{M}_{\pm} := \frac{1}{2} \( \bm{M} \pm \bm{M}^{\star} \) \, ,
\eeq
which can be alternatively stated as
\beq
\[ \,\star, \bm{M}_+ \] \equiv 0 \, , \hspace{1cm} \{ \star, \bm{M}_- \} \equiv 0 \, ,
\eeq
respectively. We also note that, as in the case of antisymmetric matrices $\bm{A}$, the odd ones $\bm{M}_-$ have no invariant component $\bra \bm{M}_- \ket \equiv \bra \star\, \bm{M}_- \ket \equiv 0$. Expressing this conjugation in terms of Lorentz indices we find
\beq
(\bm{M}^{\star})_{abcd} \equiv M_{cdab} - \et_{ca} \hat{M}_{db} + \et_{cb} \hat{M}_{da} + \et_{da} \hat{M}_{cb} - \et_{db} \hat{M}_{ca}   \, ,
\eeq
where
\beq
\hat{M}_{ab} := M^c_{\,\,\,acb} - \frac{1}{4}\, \et_{ab}\, M^{ab}_{\,\,\,\,\,\,ab} \, ,
\eeq
is identically traceless. Therefore, for a symmetric matrix $\bm{S}$
\bea
(\what{\bm{S}}_+)_{abcd} & \equiv & S_{abcd} + \frac{1}{6}\, \vep_{abcd}\, \ti{S} - \et_{a[c} \hat{S}_{d]b} - \et_{b[c} \hat{S}_{d]a}   \\
 & \equiv & \frac{1}{3} \[ S_{abcd} + \frac{1}{2} \( S_{acbd} - S_{adbc} - S_{bcad} + S_{bdac} \) + S_{cdab} \] - \et_{a[c} \hat{S}_{d]b} - \et_{b[c} \hat{S}_{d]a} \, , \nn
\eea 
is the fully traceless component $(\what{\bm{S}}_+)^c_{\,\,\,acb} \equiv 0$ with the symmetries of the Weyl tensor, while 
\beq
(\what{\bm{S}}_-)_{abcd} \equiv (\bm{S}_-)_{abcd} \equiv \et_{a[c} \hat{S}_{d]b} - \et_{b[c} \hat{S}_{d]a} \, ,
\eeq
is only partially traceless $(\bm{S}_-)^c_{\,\,\,acb} \equiv \hat{S}_{ab}$, therefore capturing the 2-tensor part of $S_{abcd}$. These roles are switched for antisymmetric matrices, i.e. the fully traceless part is the odd component $\bm{A}_-$ 
\beq
(\bm{A}_-)_{abcd} \equiv A_{abcd} + \et_{a[c} \hat{A}_{d]b} + \et_{b[c} \hat{A}_{d]a} \hspace{1cm} \Rightarrow \hspace{1cm} (\bm{A}_-)^c_{\,\,\,acb} \equiv 0 \, ,
\eeq
while the even component $\bm{A}_+$ is the partially traceless one
\beq
(\bm{A}_+)_{abcd} \equiv -\et_{a[c} \hat{A}_{d]b} + \et_{b[c} \hat{A}_{d]a} \hspace{1cm} \Rightarrow \hspace{1cm} (\bm{A}_+)^c_{\,\,\,acb} \equiv \hat{A}_{ab} \, .
\eeq
Thus, a generic matrix $\bm{M}$ can be uniquely decomposed as
\beq \label{eq:fulldecomp}
\bm{M} \equiv \frac{1}{6} \( S \, \bm{1} - \ti{S} \, \star \) + \what{\bm{S}}_+ + \bm{S}_- + \bm{A}_+ + \bm{A}_-  \, ,
\eeq 
with every term having a definite behavior under both transposition and conjugation and being normal to all others with respect to the matrix inner product \eqref{eq:inoutprods}. Moreover, notice that each of these components corresponds to an irreducible representation of the Lorentz group. In contrast to the usual definitions involving index symmetries, we see that in four dimensions the presence of $\star$ allows one to define these components in a more algebraic way, namely, by distinguishing among all the possible behaviors under matrix trace, transposition and conjugation. 

Finally, the (co-)vectors and matrices described above will always be differential forms of even degree, and it is understood that their products will always be wedge products, so their matrix/vector components commute. Also, when manipulating the Lagrangian, we will sometimes consider rational powers of 4-forms, as in \eqref{eq:SPC} for instance. These should therefore be interpreted as powers of the corresponding weight one densities, with the $\ed^4 x$ factors always appearing with the correct power at the end to make the result coordinate-independent.

\section{Plebanski actions} \label{sec:Plebanski}

Here by ``Plebanski Lagrangian" we will mean a member of the following family
\beq \label{eq:PlebanskiGen}
L = \bra B | \bm{z}_1 | F \ket - \frac{1}{2}\, \bra B | \( \bm{z}_2 + \bm{z}_3 \bm{\psi} \bm{z}_3 \) | B \ket + \ph \[ \bra \bm{z}_4 \bm{\psi} \ket - \al \] \, ,
\eeq
where 
\beq
F^{ab} := \ed A^{ab} + A^a_{\,\,\,c} \we A^{cb} \, ,
\eeq
are the curvature 2-forms of a real spin connection 1-form $A^{ab}$, while $\psi_{abcd}$, $B_{ab}$ and $\ph$ are real auxiliary 0, 2, and 4-forms, respectively, and $\bm{\psi}$ is symmetric. Moreover, the $\bm{z}_k$ are real invariant matrices with constant coefficients. Out of the nine parameters present in \eqref{eq:PlebanskiGen}, only two are relevant, because seven of them can be eliminated by linear/affine redefinitions of the auxiliary fields
\beq \label{eq:PlebRedef}
\ph \to \al' \ph \, , \hspace{1cm}  | B \ket \to \bm{z}'_1 | B \ket \, , \hspace{1cm} \bm{\psi} \to \bm{z}'_2 + \bm{z}'_3 \bm{\psi} \bm{z}'_3 \, ,
\eeq
and these relate all the possible formulations found in the literature \cite{DPF1998,CMPR2001, FS2012, SS2009, MV2010}. The two relevant parameters are ultimately $\la := 8\pi G \La$ and $\ga$ that were mentioned in the introduction. Observe, however, that the equivalence of all these Lagrangians is a feature of the Lorentzian case only, because in the Euclidean case, where $\star^2 \equiv \bm{1}$, not all real $\bm{z} \neq 0$ are invertible. Note also that the constraint imposed by $\ph$ on $\bm{\psi}$ is sometimes considered implicitly, i.e. without including the corresponding term in the Lagrangian.

\section{Derivation of the pure connection action} \label{sec:PC}

We begin with the vierbein Hilbert-Palatini-Holst Lagrangian 4-form of vacuum GR with mostly-pluses signature
\beq \label{eq:Palatini}
L = \frac{1}{8\pi G} \[ \bra E | \( \frac{\bm{1}}{\ga} + \star \) | F \ket - \La\, e \]  \, ,
\eeq
where 
\beq 
E^{ab} := e^a \we e^b \, , \hspace{1cm} e := \frac{1}{6}\, \bra E | \star | E \ket \equiv \frac{1}{4!}\,\vep_{abcd} \, e^a \we e^b \we e^c \we e^d \, , 
\eeq
$e^a$ are the vierbein 1-forms and $A^{ab}$ are the (independent) spin connection 1-forms. Keep in mind that it is also possible to include the topological terms $\bra F | \bm{w} | F \ket$, where $\bm{w}$ is some constant invariant matrix, which will only affect the quantum theory, but we neglect these here. 

The case $|\ga| \to \infty$ corresponds to the standard Hilbert-Palatini action, which is parity-symmetric (``non-chiral"), while a finite $\ga$ leads to the inclusion of the parity-violating (``chiral") Holst term \cite{Holst1995}. As in the metric case, by integrating out $A^{ab}$ one recovers the Einstein-Hilbert action for the metric $g := \et_{ab}\, e^a \otimes e^b$. In this process, the relation between the vierbein and the spin connection is independent of $\ga$ and the Holst term $\sim \bra E | F \ket$ vanishes identically, so this extension does not influence the classical dynamics. In fact, if the theory is quantized in a Lorentz-covariant manner, then the physical observables seem to be independent of $\ga$ in the quantum theory too \cite{Alexandrov2000, Livine2006}. This was definitely settled in \cite{MRC2017}, where it was shown that with the appropriate parametrization of phase space the $\ga$ dependence drops already at the level of the canonical action. 

The theory \eqref{eq:Palatini} can be obtained as a sector of the Plebanski theory \eqref{eq:PlebanskiGen} and here a convenient choice will be 
\beq \label{eq:PlebanskiPC}
L = \bra B | \bm{z} | F \ket - \frac{1}{2}\, \bra B | \bm{\psi} | B \ket + \ph \[ \bra \bm{\psi} \ket - \la \]    \, , \hspace{1cm} \la := 8\pi G \La \, ,
\eeq
where 
\beq \label{eq:zdef}
\bm{z}(\ga) := \frac{\( \ga + 1 \) \bm{1} + \( \ga - 1 \) \star}{2\ga} \, . 
\eeq 
The relation to \eqref{eq:Palatini} is obtained through the equation of motion of $\bm{\psi}$, i.e. the ``simplicity" constraints
\beq \label{eq:BBpsi}
| B \ket \bra B | = 2\ph \bm{1} \, ,
\eeq
which are solved by either
\beq \label{eq:GRsec}
| B \ket = \pm \, \frac{\bm{1} + \star}{8\pi G} \, | E \ket \, , \hspace{1cm} \ph = \frac{e}{(8\pi G)^2} \, ,
\eeq
or
\beq \label{eq:Holstsec}
| B \ket = \pm \, \frac{\bm{1} - \star}{8\pi G} \, | E \ket \, , \hspace{1cm} \ph = \frac{-e}{(8\pi G)^2} \, ,
\eeq
for some vierbein $e^a$, where we have used
\beq
| E \ket \bra E | \equiv - \star e \, .
\eeq
The solution \eqref{eq:GRsec} with the plus sign reproduces \eqref{eq:Palatini}, while the one with the minus sign corresponds to ghost-like gravity, i.e. gravitons with negative kinetic energy. On the other hand, \eqref{eq:Holstsec} leads to the same dynamics (i.e. neglecting the Holst term) as \eqref{eq:GRsec} for finite $\ga$, but with rescaled constants
\beq \label{eq:GLaresc}
G \to \ga G \, , \hspace{1cm} \La \to -\ga \La \, . 
\eeq
In the $|\ga| \to \infty$ limit the kinetic term in the Lagrangian becomes pure Holst $\sim \bra E | F \ket$, thus corresponding to a topological theory. For later reference, we will refer to \eqref{eq:GRsec} as the ``right" sector and to \eqref{eq:Holstsec} as the ``wrong" sector, since our starting point is \eqref{eq:Palatini}. The pure connection formulation being based on \eqref{eq:PlebanskiPC}, it will necessarily contain all of these four solutions, i.e. the right and wrong sectors along with their respective sign ambiguities, and we will see how to distinguish among these later on. Note also that the $|B\ket \sim |E \ket$ relations \eqref{eq:GRsec} and \eqref{eq:Holstsec} depend on the choice of Plebanski Lagrangian, but the relation \eqref{eq:GLaresc} between the two sectors does not.

In complete analogy with the treatment of the SD case \cite{CJD1989, CJD1991, CJD1992, Peldan, CJ1992, Krasnov2011a, Krasnov2011b}, we will now integrate out $B_{ab}$, then $\psi_{abcd}$ and, if $\La \neq 0$, also $\ph$. In that procedure, the non-trivial step which was resisting completion so far \cite{Krasnov2017} is finding the solution of $\psi_{abcd}$. Plugging the solution to the equation of motion of $B_{ab}$ 
\beq \label{eq:Bsol}
| B \ket = \bm{\psi}^{-1} \bm{z} | F \ket \, ,
\eeq
back inside the Lagrangian \eqref{eq:PlebanskiPC}, we get
\beq \label{eq:Lint2}
L \to \Bra \frac{1}{2} \, \bm{\psi}^{-1} \bm{\cX}_{\bm{z}} + \ph \bm{\psi} \Ket - \la \ph \, ,
\eeq
where
\beq
\bm{\cX}_{\bm{z}} := \bm{z} \bm{\cX} \bm{z} \, ,  \hspace{1cm} \bm{\cX} := | F \ket \bra F | \, ,  
\eeq
are symmetric matrices of 4-forms. Next, given the inverse matrix variation rule 
\beq
0 \equiv \de \bm{1} \equiv \de \( \bm{\psi} \bm{\psi}^{-1} \) \equiv (\de \bm{\psi})\, \bm{\psi}^{-1} + \bm{\psi} \, \de (\bm{\psi}^{-1}) \hspace{1cm} \Rightarrow \hspace{1cm} \de (\bm{\psi}^{-1}) = - \bm{\psi}^{-1} (\de \bm{\psi})\, \bm{\psi}^{-1} \, ,
\eeq
and the cyclicity of the trace, the equation of motion of $\bm{\psi}$ is
\beq \label{eq:psieq}
2\ph \bm{\psi}^2 = \bm{\cX}_{\bm{z}} \, ,
\eeq
whose solution is therefore simply
\beq \label{eq:psisol}
\bm{\psi} = \pm \, \sqrt{\frac{\bm{\cX}_{\bm{z}}}{2\ph}} \, ,
\eeq
with the choice of sign being irrelevant for the final action. Had we considered a different Plebanski formulation \eqref{eq:PlebanskiGen}, as was the case in earlier works \cite{Krasnov2017}, the equation of motion of $\bm{\psi}$ would have taken the form
\beq \label{eq:psieq2}
\bm{\psi} \bm{Q} \bm{\psi} = \bm{P} \, ,
\eeq
for two symmetric matrices $\bm{Q}, \bm{P}$, which is therefore not solved by simply taking a square root. For instance, we could have chosen \eqref{eq:PlebanskiBF}, which is obtained from \eqref{eq:PlebanskiPC} by the redefinitions \eqref{eq:redef2}, and leads to $\bm{Q} = 2\ph \bm{z}^2$ and $\bm{P} = \bm{\cX}$. Therefore, we see that by taking full advantage of the freedom in redefining the auxiliary fields of the Plebanski Lagrangian, we can considerably simplify some algebraic manipulations. Nevertheless, it is instructive to note that one can actually proceed in the general case too, i.e. that \eqref{eq:psieq2} also admits a closed-form symmetric solution 
\beq \label{eq:psisol2}
\bm{\psi} = \pm \bm{P}^{1/2} \( \bm{P}^{1/2} \bm{Q} \bm{P}^{1/2} \)^{-1/2} \bm{P}^{1/2} \, .
\eeq
In order to manipulate such expressions back inside the Lagrangian, however, it is useful to have in mind that (at least for invertible matrices) the cyclicity of the trace holds even in the presence of a matrix square root, because analytic matrix functions commute with similarity transformations
\beq \label{eq:Trsqrtcyc}
\Bra \sqrt{\bm{M} \bm{M}'} \Ket \equiv \Bra \bm{M} \bm{M}^{-1} \sqrt{\bm{M} \bm{M}'} \Ket \equiv \Bra \bm{M}^{-1} \sqrt{\bm{M} \bm{M}'} \bm{M} \Ket \equiv \Bra \sqrt{\bm{M}^{-1} \bm{M} \bm{M}' \bm{M}} \Ket \equiv \Bra \sqrt{\bm{M}' \bm{M}} \Ket \, .
\eeq
Whichever the approach we choose, the resulting Lagrangian reads
\beq \label{eq:LofXpsi}
L \to \pm \, \sqrt{2\ph} \Bra \sqrt{\bm{\cX}_{\bm{z}}} \Ket - \la \ph \, , 
\eeq 
since it cannot depend on redefinitions of fields that have been integrated out. There are now two options: either $\la = 0$ or $\la \neq 0$. In the former case $\ph$ cannot be integrated out because it is not present in its equation of motion, i.e. it acts as a Lagrange multiplier enforcing the constraint
\beq \label{eq:la0cond}
\Bra \sqrt{\bm{\cX}_{\bm{z}}} \Ket = 0 \, .
\eeq
Moreover, the equation of motion of $A^{ab}$ involves the inverse matrix $\sim \bm{\cX}_{\bm{z}}^{-1/2}$, meaning that it requires $\bm{\cX}_{\bm{z}}$ to be invertible. However, in the absence of a cosmological constant and sources, the maximally symmetric solution is Minkowski space-time $F^{ab} = 0$, meaning $\bm{\cX}_{\bm{z}} = 0$, so one cannot describe fluctuations around that solution. At the level of the action, this issue manifests itself as the fact that the square root is not differentiable at zero. In conclusion, contrary to the SD case\footnote{In the SD case with $\la = 0$ one can avoid this problem thanks to the lower dimensionality of the involved matrices \cite{CJD1991}. The Lagrangian is formally \eqref{eq:Lint2}, but the corresponding matrices are $3 \times 3$ complex symmetric. Thanks to this, by manipulating the characteristic equation of $\bm{\psi}^{-1}$ one can show that $\bra \bm{\psi} \ket \equiv \[ \bra \bm{\psi}^{-1} \ket^2 - \bra \bm{\psi}^{-2} \ket \]/2 \det \bm{\psi}$, which makes \eqref{eq:Lint2} quadratic in $\bm{\psi}^{-1}$, after a redefinition of $\ph$ to absorb the determinant. Thus, integrating out $\bm{\psi}^{-1}$ leads to a Lagrangian of the form \eqref{eq:LofXpsi} with $\la = 0$, but with the analogue of $\bm{\cX}_{\bm{z}}$ entering quadratically. Unfortunately, in the real case this trick does not work because the characteristic polynomial is of order six instead of three, meaning that $\bra \bm{\psi} \ket$ can be expressed in terms of $\bra \bm{\psi}^{-n} \ket$, where $n$ goes up to five. This alternative constraint can therefore only make the equation of motion of $\bm{\psi}$ worst, i.e. of higher order than the quadratic one \eqref{eq:psieq}.}, the real pure connection theory with $\la = 0$ is not well-defined in general. 

Let us next consider the $\la \neq 0$ case, where now the $\ph$ field can be integrated out of \eqref{eq:LofXpsi}. Plugging the solution to its equation of motion
\beq \label{eq:phisol}
\sqrt{2\ph} = \pm\, \frac{1}{\la} \Bra \sqrt{\bm{\cX}_{\bm{z}}} \, \Ket \, ,
\eeq
back inside \eqref{eq:LofXpsi}, we find the pure spin connection Lagrangian
\beq \label{eq:LPC}
L = \frac{1}{2\la} \Bra \sqrt{\bm{\cX}_{\bm{z}}} \, \Ket^2  \, ,
\eeq
hence the action \eqref{eq:SPC}. We can obtain an alternative form by using \eqref{eq:Trsqrtcyc} 
\beq \label{eq:LPC2}
L \equiv \frac{1}{2\la} \Bra \sqrt{\bm{z}^2\bm{\cX}}\, \Ket^2 \equiv \frac{1}{2\la} \Bra \sqrt{ \( \frac{1}{\ga}\, \bm{1} + \frac{\ga^2-1}{2\ga^2} \,  \star \) \bm{\cX}}\, \Ket^2 \, ,
\eeq
which is simpler, although the matrix argument is no longer symmetric in general. In the non-chiral case $|\ga| \to \infty$ we have
\beq \label{eq:Lspec}
\lim_{\ga \to \pm \infty} L = \frac{1}{4\la} \Bra \sqrt{\star \bm{\cX}} \Ket^2 \, ,
\eeq
so the $\star$ combines with the Levi-Civita symbol hiding in the wedge product $F^{ab} \we F^{cb}$ to yield a parity-symmetric Lagrangian indeed. To obtain the equations of motion of \eqref{eq:LPC}, we note that the variation of an analytic matrix function inside a trace is simply $\de \bra f(\bm{M}) \ket \equiv \bra f'(\bm{M})\, \de \bm{M} \ket$, which is shown using the Taylor expansion and the cyclicity of the trace, and find 
\beq
\ced \[ \Bra \sqrt{\bm{\cX}_{\bm{z}}} \Ket \bm{\cX}_{\bm{z}}^{-1/2} \bm{z} \]_{abcd} \we F^{cd}  = 0 \, ,
\eeq
where $\ced$ is the exterior covariant derivative, we have used $[\ced ,\star] \equiv 0$ and the Bianchi identity $\ced F^{ab} \equiv 0$. Now the presence of the inverse of $\bm{\cX}_{\bm{z}}$ is no longer a problem, because the maximally symmetric solution is (A)dS, for which $\bm{\cX} \sim \star$ is invertible, so we do have access to the dynamics of the corresponding fluctuations. More generally, we can describe any space-time for which $\bm{\cX}$ is invertible everywhere, since $\bm{z}$ is also invertible for real $\ga$. 

Let us finally come back to the issue of selecting the set of solutions corresponding to the original theory \eqref{eq:Palatini}. As we will see in sections \ref{sec:chiral} and \ref{sec:linear}, it will sometimes be possible to distinguish between the ``right" and ``wrong" sectors, but the action will be insensitive to the choice of sign within each sector. This is because the two options are related by the sign flip $| F \ket \to -| F \ket$ at the level of the Hilbert-Palatini-Holst Lagrangian \eqref{eq:Palatini}, and the pure connection one \eqref{eq:LPC} only depends on the quadratic combination $\bm{\cX} := | F \ket \bra F |$, which is invariant. Nevertheless, this will not be a problem for practical purposes. Indeed, given the relation \eqref{eq:Bsol} between $| B \ket$ and $| F \ket$, we see that for $| B \ket$ to change sign $| F \ket$ must go through zero, meaning that $\bm{\cX}$ must become non-invertible, in which case the whole construct breaks down anyway. Thus, restricting to solutions where $\bm{\cX}$ is invertible everywhere automatically selects a definite sign within each sector.

\section{Derivation of the constraint-free BF action} \label{sec:BF}

Here we find it more convenient to consider as our starting point the following Plebanski Lagrangian
\beq \label{eq:PlebanskiBF}
L = \bra B | F \ket - \frac{1}{2}\, \bra B | \bm{\psi} | B \ket + \ph \[ \Bra \bm{z}^2 \bm{\psi} \Ket - \la \] \, ,
\eeq
which is obtained from \eqref{eq:PlebanskiPC} with the redefinitions
\beq \label{eq:redef2}
| B \ket \to \bm{z}^{-1} | B \ket \, , \hspace{1cm} \bm{\psi} \to \bm{z} \bm{\psi} \bm{z} \, .
\eeq
Following \cite{Krasnov2017}, we consider a redefinition of the form
\beq \label{eq:BFredef}
| B \ket \to \bm{\ch}_1 | B \ket + \bm{\ch}_2 | F \ket \, , 
\eeq
where $\bm{\ch}_{1,2}$ are such that the resulting Lagrangian maintains its ``canonical" BF form, i.e. the $\sim BF$ term is simply $\bra B | F \ket$ and the $\sim FF$ term arises only through the topological combinations $\bra F | \bm{w}^{-2} | F \ket/2$, for some constant invariant matrix $\bm{w}$, which we neglect anyway. We find it actually more convenient to proceed in two steps. First, we demand that the resulting matrices in $\bra B | \dots | F \ket$ and $\bra F | \dots | F \ket/2$ be the same invariant matrix $\bm{w}^{-2}$, so that the corresponding conditions read
\beq
\bm{\ch}_1 - \frac{1}{2} \[ \bm{\ch}_1 \bm{\psi} \bm{\ch}_2 + \bm{\ch}_2 \bm{\psi} \bm{\ch}_1 \] = \bm{w}^{-2} \, , \hspace{1cm} \bm{\ch}_2 - \frac{1}{2}\, \bm{\ch}_2 \bm{\psi} \bm{\ch}_2 = \frac{1}{2}\, \bm{w}^{-2} \, ,
\eeq
and admit the symmetric solutions
\beq
\bm{\ch}_1 = \bm{w}^{-1} \frac{\pm \bm{1}}{\sqrt{\bm{1} - \bm{\psi}_{\bm{w}}}} \, \bm{w}^{-1} \, , \hspace{1cm} \bm{\ch}_2 = \bm{w}^{-1} \frac{\bm{1} \mp \sqrt{\bm{1} - \bm{\psi}_{\bm{w}}}}{\bm{\psi}_{\bm{w}}} \, \bm{w}^{-1} \, , 
\eeq
where
\beq
\bm{\psi}_{\bm{w}} := \bm{w}^{-1} \bm{\psi} \bm{w}^{-1} \, .
\eeq
The resulting Lagrangian reads
\beq \label{eq:LBFpsi0}
L \to \bra B | \bm{w}^{-2} | F \ket - \frac{1}{2}\, \bra B | \bm{w}^{-1} \frac{\bm{\psi}_{\bm{w}}}{\bm{1} - \bm{\psi}_{\bm{w}}}\, \bm{w}^{-1} | B \ket + \ph \[ \bra \bm{z}^2 \bm{\psi} \ket - \la \] \, ,
\eeq 
so we perform one more redefinition
\beq
| B \ket \to \bm{w}^2 | B \ket \, , \hspace{1cm} \bm{\psi} \to \bm{w} \, \frac{\bm{\psi}_{\bm{w}}}{\bm{1} + \bm{\psi}_{\bm{w}}}\, \bm{w} \, ,
\eeq
to obtain the canonical BF form
\beq \label{eq:LBFpsi}
L \to \bra B | F \ket - \frac{1}{2}\, \bra B | \bm{\psi} | B \ket + \ph \[ \Bra \( \bm{z}\bm{w} \)^2 \frac{\bm{\psi}_{\bm{w}}}{\bm{1} + \bm{\psi}_{\bm{w}}} \Ket - \la \] \, .
\eeq 
The Plebanski Lagrangian now corresponds to the $\bm{w}^{-1} \to \bm{0}$ limit. For finite $\bm{w}$, however, \eqref{eq:LBFpsi} is no longer linear in $\bm{\psi}$, so the latter can be integrated out without constraining other fields. Its equation of motion can be put in the form \eqref{eq:psieq2}
\beq
\( \bm{1} + \bm{\psi}_{\bm{w}} \) \bm{w} \bm{\cY} \bm{w} \( \bm{1} + \bm{\psi}_{\bm{w}} \) = 2\ph \( \bm{z} \bm{w} \)^2 \, , \hspace{1cm} \bm{\cY} := | B \ket \bra B | \, ,
\eeq
and is therefore solved using \eqref{eq:psisol2}
\beq
\bm{\psi} = \bm{w} \[ \pm \sqrt{2\ph}\, \bm{z} \bm{w} \( \bm{z} \bm{w}^2 \bm{\cY} \bm{w}^2 \bm{z} \)^{-1/2} \bm{w} \bm{z} - \bm{1} \] \bm{w} \, ,
\eeq
leading to
\beq
L \to \bra B | F \ket + \frac{1}{2} \Bra \bm{w} \bm{\cY} \bm{w} \Ket \mp \sqrt{2\ph} \Bra \sqrt{\bm{z} \bm{w}^2 \bm{\cY} \bm{w}^2 \bm{z}} \,\Ket + \ph \[ \Bra \( \bm{z} \bm{w} \)^2 \Ket - \la \] \, .
\eeq
Finally, integrating out $\ph$, whose solution reads
\beq
\sqrt{2\ph} = \pm \, \frac{\Bra \sqrt{\bm{z} \bm{w}^2 \bm{\cY} \bm{w}^2 \bm{z}}\, \Ket}{\Bra \( \bm{z} \bm{w} \)^2 \Ket - \la} \, ,
\eeq
we get the constraint-free BF Lagrangian
\beq \label{eq:LBF}
L \to \bra B | F \ket + \frac{1}{2} \[ \Bra \bm{w} \bm{\cY} \bm{w} \Ket + \frac{\Bra \sqrt{\bm{z} \bm{w}^2 \bm{\cY} \bm{w}^2 \bm{z}}\, \Ket^2}{\la - \Bra \( \bm{z} \bm{w} \)^2 \Ket} \] \, ,
\eeq
hence the action \eqref{eq:SBF}. As in the pure connection case, we can use \eqref{eq:Trsqrtcyc} to write this as
\beq  \label{eq:LBF2}
L \equiv \frac{1}{2} \[ B_{ab} \we F^{ab} + \Bra \bm{w}^2 \bm{\cY} \Ket + \frac{\Bra \sqrt{\bm{z}^2 \bm{w}^4 \bm{\cY}}\, \Ket^2}{\la - \Bra \( \bm{z} \bm{w} \)^2 \Ket} \] \, ,
\eeq
at the expense of having non-symmetric matrix arguments. Note that, contrary to the pure connection case, here we do have access to the $\la = 0$ theory. We still have a matrix square root leading to inverse matrices in the equations of motion, but now this concerns $\bm{\cY}$, which is invertible on the Minkowski solution because $B^{ab}$ is essentially the vierbein. 

As a last check of the pure connection Lagrangian \eqref{eq:LPC}, we can now also integrate out $| B \ket$. Its equation of motion reads 
\beq \label{eq:BeqBF}
\[ \bm{w}^2 + \frac{\Bra \sqrt{\bm{z} \bm{w}^2 \bm{\cY} \bm{w}^2 \bm{z}}\, \Ket}{\la - \Bra \( \bm{z} \bm{w} \)^2 \Ket} \, \bm{z} \bm{w}^2 \( \bm{z} \bm{w}^2 \bm{\cY} \bm{w}^2 \bm{z} \)^{-1/2} \bm{w}^2 \bm{z} \] | B \ket = -| F \ket \, ,
\eeq
so we must express $\bm{\cY}$ in the square bracket in terms of $\bm{\cX} \equiv | F \ket \bra F |$ in order to invert this. We first take the outer product of each side with itself, multiply by $\bm{z}$ from right and left to form a total square and then take the square root
\beq \label{eq:XzofB}
\sqrt{\bm{\cX}_{\bm{z}}} = \pm \[ \sqrt{\bm{z} \bm{w}^2 \bm{\cY} \bm{w}^2 \bm{z}} + \( \bm{z} \bm{w} \)^2 \frac{\Bra \sqrt{\bm{z} \bm{w}^2 \bm{\cY} \bm{w}^2 \bm{z}}\, \Ket}{\la - \Bra \( \bm{z} \bm{w} \)^2 \Ket}  \] \, .
\eeq
Next, taking the trace of this
\beq
\Bra \sqrt{\bm{\cX}_{\bm{z}}}\, \Ket = \pm \, \frac{\la \Bra \sqrt{\bm{z} \bm{w}^2 \bm{\cY} \bm{w}^2 \bm{z}} \, \Ket}{\la - \Bra \( \bm{z} \bm{w} \)^2 \Ket}  \, .
\eeq
and plugging the result back inside \eqref{eq:XzofB}, we obtain 
\beq \label{eq:shortcut1}
\sqrt{\bm{z} \bm{w}^2 \bm{\cY} \bm{w}^2 \bm{z}} = \pm \[ \sqrt{\bm{\cX}_{\bm{z}}} - \frac{1}{\la} \Bra \sqrt{\bm{\cX}_{\bm{z}}}\, \Ket \( \bm{z} \bm{w} \)^2 \] \, ,
\eeq
which, once plugged inside \eqref{eq:BeqBF}, allows us to isolate $| B \ket$
\beq
| B \ket = \bm{w}^{-1}\, \frac{\Bra \sqrt{\bm{\cX}_{\bm{z}}} \, \Ket \bm{1} - \la \( \bm{z} \bm{w} \)^{-1} \sqrt{\bm{\cX}_{\bm{z}}} \( \bm{z} \bm{w} \)^{-1}}{\la \( \bm{z} \bm{w} \)^{-1}\sqrt{\bm{\cX}_{\bm{z}}}\( \bm{z} \bm{w} \)^{-1}} \,\bm{w}^{-1} | F \ket \, .
\eeq
Finally, plugging this back inside \eqref{eq:LBF}, and using also the shortcut \eqref{eq:shortcut1}, we recover the pure connection Lagrangian \eqref{eq:LPC}, up to topological terms $\sim F^2$. The latter are $-\bra \bm{w}^{-1} \bm{\cX} \bm{w}^{-1} \ket/2$, thus precisely canceling the $\bra F | \bm{w}^{-2} | F \ket/2$ term that was introduced by the redefinition \eqref{eq:BFredef}. The resulting pure connection Lagrangian is therefore consistently independent of $\bm{w}$, since the latter controls the redefinition of a field that has been integrated out.

\section{The SL$(2,\Cs)$ and SD formulations} \label{sec:chiral}

Let us now express the Lagrangians \eqref{eq:LPC} and \eqref{eq:LBF} in terms of the universal covering group SL$(2,\Cs)$, in which case the independent variables are the self-dual components
\bea
A^i & := & -\frac{1}{2} \, \vep^{ijk} A^{jk} + i A^{0i} \, ,  \\
F^i & := & -\frac{1}{2}\, \vep^{ijk} F^{jk} + i F^{0i} \equiv \ed A^i + \frac{1}{2}\, \vep^{ijk} A^j \we A^k \, , \\
B^i & := & \frac{1}{2} \[ -\frac{1}{2}\,\vep^{ijk} B^{jk} + i B^{0i} \]  \, ,
\eea
and their complex conjugates, and we use $i,j,k,\dots$ to denote the spatial Lorentz indices that are displaced with $\de^{ij}$. In terms of these quantities the Lagrangians are formally the same as \eqref{eq:LPC} and \eqref{eq:LBF}, only now
\beq \label{eq:vepXC}
\star := \( \begin{array}{cc} i \de^{ij} & 0 \\ 0 & - i \de^{ij} \end{array} \) \, , \hspace{0.7cm} \bm{\cX} := \frac{1}{2} \( \begin{array}{cc} F^i \we F^j & F^i \we \bar{F}^j \\ \bar{F}^i \we F^j & \bar{F}^i \we \bar{F}^j \end{array} \) \, ,  \hspace{0.7cm} \bm{\cY} := 2 \( \begin{array}{cc} B^i \we B^j & B^i \we \bar{B}^j \\ \bar{B}^i \we B^j & \bar{B}^i \we \bar{B}^j \end{array} \) \, , 
\eeq
which are $6 \times 6$ complex symmetric matrices. In particular, a real invariant matrix $\bm{z}$ becomes
\beq \label{eq:vepXC2}
\bm{z} \equiv \( \begin{array}{cc} z \de^{ij} & \bm{0} \\ \bm{0} & \bar{z} \de^{ij} \end{array} \) \, , \hspace{1cm} z := \bm{z}|_{\bm{1} \to 1,\star \to i} \, .
\eeq
To prove \eqref{eq:vepXC} one must simply check that all the traces $\{ \bra \bm{\cX}_{\bm{z}}^n \ket \}_{n = 1, \dots, 6}$ coincide, because these numbers fully determine the characteristic polynomial of $\bm{\cX}_{\bm{z}}$, and thus its eigenvalues, which in turn fully determine $L$. The same then holds for the functions of $\bm{\cY}$. With the SL$(2,\Cs)$ forms at hand, we can now recover the SD Lagrangians in their pure-connection and BF forms. We start with the former, which is given by \cite{Krasnov2011a, Krasnov2011b}
\bea \label{eq:LK}
L_{\rm SD}[F] := \frac{i}{2\la} \Bra \sqrt{\bm{X}} \, \Ket^2  \, , \hspace{1cm} X^{ij} := F^i \we F^j \, ,
\eea
and is supplemented by the reality constraints \cite{HKS2016}
\beq \label{eq:realconstrF}
F^i \we \bar{F}^j = 0 \, , \hspace{1cm} {\rm Re} \[ \Bra \sqrt{\bm{X}} \Ket^2 \] = 0 \, ,
\eeq
with the latter making the Lagrangian real. Note that the complex metric defined by \cite{Urbantke,HKS2016}
\beq \label{eq:UrbGR}
g_{\al\be} \sim \vep_{ijk} \, \vep^{\mu\nu\ro\si} F^i_{\al\mu} F^j_{\nu\ro} F^k_{\si\be} \, , \hspace{1cm} \sqrt{|g|} \, \ed^4 x = \frac{\la}{\La^2}\, L_{\rm SD}  \, ,
\eeq
is the one satisfying the Einstein equations when $A^i$ is a solution \cite{HKS2016}, and the ten reality constraints \eqref{eq:realconstrF} are equivalent to the statement that this metric is real Lorentzian. Since this property is preserved under evolution, these constraints are compatible with the dynamics from the canonical viewpoint. 

The easiest way to obtain \eqref{eq:LK} from the real Lagrangian is to simply set $\ga = -i$ in \eqref{eq:LPC2} with \eqref{eq:vepXC}, thus projecting out the anti-self-dual component 
\beq
\bm{z}^2 \to \( \begin{array}{cc} 2 i \de^{ij} & \bm{0} \\ \bm{0} & \bm{0} \end{array} \) \hspace{1cm} \Rightarrow \hspace{1cm} L \to \frac{i}{2\la}\, \Bra \sqrt{\bm{X}} \Ket^2 \, .
\eeq  
Another option is to impose the reality constraints \eqref{eq:realconstrF} on \eqref{eq:LPC2} using \eqref{eq:vepXC2} for generic $\ga \in \Rs$. The first constraint of \eqref{eq:realconstrF} makes $\bm{\cX}$ block-diagonal, so 
\beq
\sqrt{\bm{z}^2 \bm{\cX}} \to \frac{1}{\sqrt{2}} \( \begin{array}{cc} z \sqrt{\bm{X}} & \bm{0} \\ \bm{0} & \overline{z\sqrt{\bm{X}}} \end{array} \) \hspace{1cm} \Rightarrow \hspace{1cm} L \to \frac{1}{4\la} \Bra z \sqrt{\bm{X}} + \overline{z \sqrt{\bm{X}}} \Ket^2 \, .
\eeq
Then, rewriting the second constraint of \eqref{eq:realconstrF} as
\beq \label{eq:scalconstr}
{\rm Re} \Bra \sqrt{\bm{X}} \Ket = \pm {\rm Im} \Bra \sqrt{\bm{X}} \Ket \hspace{1cm} \Rightarrow \hspace{1cm} \Bra \sqrt{\bm{X}} \Ket \pm i \Bra \overline{\sqrt{\bm{X}}} \Ket = 0 \, ,
\eeq
we find, depending on the choice of sign,
\beq
L_+ \to \frac{i}{2\la} \Bra \sqrt{\bm{X}} \Ket^2  \, , \hspace{1cm} L_- \to \frac{-i}{2\ga^2\la} \Bra \sqrt{\bm{X}} \Ket^2 \, .
\eeq
Thus, with the plus sign we recover \eqref{eq:LK} indeed, while with the minus sign the constants are rescaled as in \eqref{eq:GLaresc}, meaning that this corresponds to the wrong sector of the theory. It is therefore more accurate to state \eqref{eq:scalconstr} with ``$+$" as the scalar constraint leading to GR, rather than the one in \eqref{eq:realconstrF}.

We next want to derive the BF form without simplicity constraints in the SD case. Note that, contrary to the SD Plebanski Lagrangian, where the reality constraints are expressed exclusively in terms of $B^i$, here this is no longer the case, because that field has been mixed with $F^i$ in the redefinition
\beq
B^i \to \chi_1^{ij} B^j + \chi_2^{ij} F^j \, .
\eeq
For this reason, here we will only consider the easier approach which is to simply set $\ga = -i$. As in the pure-connection Lagrangian, the resulting $\bm{z}^2$ projects out all $\bar{B}^i$ dependencies in the matrix square root term of \eqref{eq:LBF2}, while in the other two terms the anti-self-dual components are already decoupled. We can therefore collect the SD sector to find
\beq \label{eq:LKBF}
L_{\rm SD}[B,F] := B^i \we F^i + w^2 \[ \Bra \bm{Y} \Ket + \frac{2i w^2}{\la - 6i w^2} \Bra \sqrt{\bm{Y}} \, \Ket^2  \] \, , \hspace{1cm} Y^{ij} := B^i \we B^j \, ,
\eeq
where $w$ is the complex number in $\bm{w}$ according to the relation \eqref{eq:vepXC2}. The above Lagrangian is indeed the Lorentzian version of the one found in \cite{HK2015,Krasnov2017} which is obtained by sending $\la \to i \la$.

\section{The linearized pure connection action over (A)dS}  \label{sec:linear}

Let us now consider the (A)dS solution $\bar{A}^{ab}$, in which case there exists some vierbein $\bar{e}^a$ such that
\beq \label{eq:AdSAF}
\bar{\ced} \bar{e}^a \equiv \ed \bar{e}^a + \bar{A}^a_{\,\,\,b} \we \bar{e}^b = 0 \, , \hspace{1cm} \bar{F}^{ab} = \frac{\al\La}{3} \, \bar{e}^a \we \bar{e}^b   \, ,
\eeq
where $\al = 1$ for the solution in the right sector and $\al = - \ga$ in the wrong one, given the relation \eqref{eq:GLaresc} between the two. In particular, the matrix entering the Lagrangian \eqref{eq:LPC} is
\beq \label{eq:AdSpt}
\bar{\bm{\cX}}_{\bm{z}} = -\frac{\al^2\La^2 \bar{e}}{9}\, \bm{z}^2 \star \equiv \frac{\al^2\La^2 \bar{e}}{9} \( \frac{\ga^2 - 1}{2\ga^2}\, \bm{1} - \frac{1}{\ga}\, \star \)   \, .
\eeq
We next consider a perturbation around that solution 
\beq
\de A^{ab} := A^{ab} - \bar{A}^{ab} \, , \hspace{1cm} \de F^{ab} := F^{ab} - \bar{F}^{ab} \equiv \bar{\ced}\, \de A^{ab} + \de A^a_{\,\,\,c} \we \de A^{cb} \, ,
\eeq
and compute the part of the Lagrangian that is second order in $\de A^{ab}$, i.e. the linearized theory. In order to expand the Lagrangian in powers of $\de A^{ab}$, we need to expand the trace of a matrix square root around $\bar{\bm{\cX}}_{\bm{z}}$. However, the latter is not a multiple of the identity when $\ga$ is finite, so we cannot simply Taylor expand without caring about the order of the matrices. We therefore require a bit more algebra before we can begin. We first decompose $\bm{\cX}_{\bm{z}}$ into irreducible pieces \eqref{eq:fulldecomp}
\beq
\bm{\cX}_{\bm{z}} \equiv \frac{1}{6} \( S \bm{1} - \ti{S}\, \star \) + \what{\bm{S}}_+ + \bm{S}_- \, ,
\eeq
and note that the last two are first order in $\de A^{ab}$, since the background \eqref{eq:AdSpt} is an invariant matrix, so we can expand the Lagrangian to second order in $\what{\bm{S}}_+$ and $\bm{S}_-$. The desired expression is found by taking the most general symmetric ansatz containing the involved matrices
\bea
\sqrt{\bm{\cX}_{\bm{z}}} & = & a \bm{1} + \ti{a} \star + \,a_+ \what{\bm{S}}_+ + \ti{a}_+ \star \what{\bm{S}}_+ + a_-\, \bm{S}_- + a_{++} \, \what{\bm{S}}_+^2 + \ti{a}_{++} \star \what{\bm{S}}_+^2 +  \nn \\
 & & +\, a_{--} \, \bm{S}_-^2 + \ti{a}_{--} \star \bm{S}^2_- + a_{+-} \( \what{\bm{S}}_+ \bm{S}_- + \bm{S}_- \what{\bm{S}}_+ \) + \Ord(\what{\bm{S}}_{\pm}^3) \, , 
\eea
squaring both sides and solving for the coefficients, which therefore depend on $S$ and $\ti{S}$. Taking the trace of the result and then the square, only four combinations  survive
\beq \label{eq:Lagexp}
\Bra \sqrt{\bm{\cX}_{\bm{z}}} \Ket^2 \equiv c - \frac{9}{2} \Bra \bm{c}_+ \what{\bm{S}}_+^2 + c_- \bm{S}_-^2 \Ket + \Ord(\what{\bm{S}}^3) \, .
\eeq
Indeed, all first order terms drop thanks to the tracelessness of $\what{\bm{S}}_{\pm}$, $\bm{S}_{\pm} \bm{S}_{\mp}$ and $\star \bm{S}_{\pm} \bm{S}_{\mp}$ are odd, so their trace vanishes too, and we have used the cyclicity of the trace to also get 
\beq
\Bra \star\, \bm{S}_-^2 \Ket \equiv - \Bra \bm{S}_- \star \bm{S}_- \Ket \equiv - \Bra \star\, \bm{S}_-^2 \Ket  \hspace{1cm} \Rightarrow \hspace{1cm} \Bra \star\, \bm{S}_-^2 \Ket \equiv 0 \, .
\eeq
The coefficients $c$, $\bm{c}_+$ and $c_-$ in \eqref{eq:Lagexp} are determined up to a sign ambiguity, e.g.
\beq \label{eq:cofSSt}
c = 3 \( S \pm \sqrt{S^2 + \ti{S}^2} \) \, ,
\eeq
which, in fact, distinguishes the right and wrong sectors. To determine which is which, we first evaluate the Lagrangian on the (A)dS solution \eqref{eq:AdSpt} to find
\beq \label{eq:LofbarApm}
L[\bar{A}] = \frac{\bar{c}}{16\pi G \La} \equiv \frac{\al^2\La \bar{e}}{16\pi G} \frac{\ga^2 - 1 \pm \( \ga^2 + 1 \)}{\ga^2} \, ,
\eeq 
and compare with the original Lagrangian \eqref{eq:Palatini} evaluated on the same solution \eqref{eq:AdSAF}
\beq \label{eq:HPHonAds}
L[\bar{e}, \bar{A}] = \frac{\La \bar{e}}{8\pi G} \, .
\eeq
We see that the right sector $\al = 1$ corresponds to the plus sign in \eqref{eq:LofbarApm}, whereas the wrong sector $\al = -\ga$, which is related to \eqref{eq:HPHonAds} by the replacement \eqref{eq:GLaresc}, is consistently reproduced by taking the minus sign in \eqref{eq:LofbarApm}. Focusing on the right sector from now on, the coefficients in \eqref{eq:Lagexp} are given by
\beq \label{eq:ccoefp}
c = 3 \( S + \sqrt{S^2 + \ti{S}^2} \) \, , \hspace{0.5cm} \bm{c}_+ = \frac{\( S\, \bm{1} + \ti{S} \,\star \)^2 + \( S\, \bm{1} + \ti{S}\, \star \) \sqrt{S^2 + \ti{S}^2}}{\( S^2 + \ti{S}^2 \)^{3/2}}  \, ,  \hspace{0.5cm} c_- = \frac{2}{\sqrt{S^2 + \ti{S}^2}} \, . 
\eeq
As far as the $\Ord(\de A^2)$ part of the Lagrangian is concerned, $\bm{c}_+$ and $c_-$ can be directly evaluated on the background solution \eqref{eq:AdSpt}
\beq \label{eq:bmu}
\bar{\bm{c}}_+ \equiv \frac{6\ga^3}{\La^2 \bar{e} \( \ga^2 + 1 \)^3} \[ \ga \( \ga^2 - 3 \) \bm{1} + \( 3 \ga^2 - 1 \) \star \]  \, ,  \hspace{1cm} \bar{c}_- \equiv \frac{6\ga^2}{\La^2 \bar{e} \( \ga^2 + 1 \)} \, , 
\eeq
since they multiply quantities that are already $\Ord(\de A^2)$. As for $c$, the most general form of the second-order part in $\de A^{ab}$ is
\beq \label{eq:mu02}
c^{(2)} \equiv \al_1 \, \bra \star \bm{\cX}^{(1)} \ket^2 + 2 \al_2 \, \bra \star \bm{\cX}^{(1)} \ket \bra \bm{\cX}^{(1)} \ket + \al_3 \, \bra \bm{\cX}^{(1)} \ket^2 + \be_1 \, \bra \star \bm{\cX}^{(2)} \ket + \be_2 \, \bra \bm{\cX}^{(2)} \ket \, ,
\eeq
but the $\sim \be_{1,2}$ terms do not contribute to the equations of motion, because they correspond to the second-order part of the topological terms $\bra F | (\be_1 \star + \,\be_2\, \bm{1}) | F \ket$ which we neglect here. It then turns out that for this particular $c(S,\ti{S})$ function \eqref{eq:ccoefp} the $\al_{1,2}$ terms vanish as well, leaving us with
\beq \label{eq:mu02}
c^{(2)} \to \frac{9}{8 \La^2 \bar{e}} \frac{\ga^2 + 1}{\ga^2} \, \bra \bm{\cX}^{(1)} \ket^2  \, .
\eeq
We thus need to compute $\bra \bm{\cX} \ket$, $\what{\bm{S}}_+$ and $\bm{S}_-$ to linear order in $\de A^{ab}$. In particular, we want to express these in terms of the components of the curvature 2-form $F^{ab}$ in the background space-time basis $\bar{e}^a$ 
\beq
\cF_{abcd} := \bar{e}_a^{\mu} \bar{e}_b^{\nu} F_{\mu\nu cd} \, , \hspace{1cm} \cF_{ab} := \cF^c_{\,\,\,acb} \, , \hspace{1cm} \cF := \cF^a_a \, ,
\eeq
so that we can now treat $\cF_{abcd}$ as a non-symmetric matrix and in particular
\beq \label{eq:XofcF}
\bar{\bm{\cF}} \equiv \frac{\La}{3} \, \bm{1} \, , \hspace{1cm} \bm{\cX} \equiv - \bm{\cF}^{\ka} \star \bm{\cF}\, \bar{e} \, .
\eeq
Perturbing this expression, and decomposing $\de \bm{\cF}$ into irreducible components \eqref{eq:fulldecomp}
\beq \label{eq:decFdecomp}
\de \bm{\cF} \equiv \frac{1}{6} \( \cS \bm{1} - \ti{\cS}\, \star \) + \what{\bm{\cS}}_+ + \bm{\cS}_- + \bm{\cA}_+ + \bm{\cA}_- \, ,
\eeq
we find
\beq \label{eq:cXpert}
\de \bm{\cX} \equiv - \frac{\La\bar{e}}{3} \[ \star\, \de \bm{\cF} + \de \bm{\cF}^{\ka} \star \] \equiv -\frac{2\La\bar{e}}{3} \[ \frac{1}{6} \( \ti{\cS} \bm{1} + \cS \, \star \) + \star \( \what{\bm{\cS}}_+ + \bm{\cA}_- \) \]  \, ,
\eeq
and thus
\beq
\de \bm{\cX}_{\bm{z}} \equiv - \frac{2\La\bar{e}}{3} \[ \frac{1}{6}\, \bm{z}^2 \( \ti{\cS} \bm{1} + \cS \, \star  \) + \star \( \bm{z}^2 \what{\bm{\cS}}_+ + | \bm{z} |^2 \bm{\cA}_- \) \]  \, , \hspace{1cm} | \bm{z} |^2 \equiv \frac{\ga^2 + 1}{2 \ga^2} \, .
\eeq
The quantities of interest are therefore
\beq
\bra \bm{\cX} \ket \equiv -\frac{2\La\bar{e}}{3}\, \ti{\cS} + \Ord(\de A^2) \, , \hspace{0.5cm} \what{\bm{S}}_+ \equiv -\frac{2\La\bar{e}}{3}\, \bm{z}^2 \star \what{\bm{\cS}}_+ + \Ord(\de A^2) \, , \hspace{0.5cm} \bm{S}_- \equiv -\frac{2\La\bar{e}}{3}\, |\bm{z}|^2 \star \bm{\cA}_- + \Ord(\de A^2) \, ,
\eeq
so plugging these inside \eqref{eq:Lagexp} and using \eqref{eq:bmu} and \eqref{eq:mu02} gives
\bea
\( \Bra \sqrt{\bm{\cX}_{\bm{z}}} \Ket^2 \)^{(2)} & \equiv & c^{(2)} + 2 \La^2 \bar{e}^2 \Bra \bar{\bm{c}}_+ \bm{z}^4 \what{\bm{\cS}}_+^2 - \bar{c}_- |\bm{z}|^4 \bm{\cA}_-^2 \Ket  \\
 & \equiv & \[ \frac{\ga^2 + 1}{2\ga^2}\, \ti{\cS}^2 - 3 \Bra \( \bm{1} - \frac{\star}{\ga} \) \what{\bm{\cS}}_+^2 + \frac{\ga^2 + 1}{\ga^2} \, \bm{\cA}_-^2 \Ket \] \bar{e}  \, ,  \nn
\eea
and we must finally compute $\ti{\cS}$, $\what{\bm{\cS}}_+$ and $\bm{\cA}_-$ in terms of $\de \bm{\cF}$ through \eqref{eq:decFdecomp}. We find
\beq
\ti{\cS} \equiv \bra \star\, \bm{\cC} \ket \, , \hspace{1cm} \what{\bm{\cS}}_+ \equiv \frac{1}{2} \[ \, \bm{\cC} + \bm{\cC}^{\ka} + \frac{1}{3}\, \bra \star\, \bm{\cC} \ket\, \star \, \] \, , \hspace{1cm} \bm{\cA}_- \equiv \frac{1}{2} \[\, \bm{\cC} - \bm{\cC}^{\ka} \] \, ,
\eeq
where
\beq
\cC_{abcd} := \cF_{abcd} - \frac{1}{2} \( \et_{ac} \cF_{bd} - \et_{bc} \cF_{ad} - \et_{ad} \cF_{bc} + \et_{bd} \cF_{ac} \) + \frac{1}{6}\, \ka_{abcd}\, \cF \, ,
\eeq
is traceless, having therefore zero background, but without the symmetries of the Weyl tensor. The final result can then be put in the following form
\beq \label{eq:SPClin}
S_{\rm (A)dS} = -\frac{3}{64 \pi G \La} \int \bar{e} \[ \frac{\ga^2 + 1}{\ga^2}\, \cC_{abcd}\, \cC^{cdab} - \frac{1}{\ga^2}\, \what{\cS}^+_{abcd}\, \what{\cS}_+^{abcd} - \frac{1}{\ga}\, \cC_{abcd}\, \ti{\cC}^{abcd} \] + \Ord(\de A^3) \, , 
\eeq
where 
\beq
\what{\cS}^+_{abcd} \equiv \frac{1}{3} \[ \cC_{abcd} + \frac{1}{2} \( \cC_{acbd} - \cC_{adbc} - \cC_{bcad} + \cC_{bdac} \) + \cC_{cdab} \] \, , \hspace{0.5cm} \ti{\cC}_{abcd} := \frac{1}{2}\, \vep_{ab}^{\,\,\,\,\,\,ef} \, \cC_{efcd} \, ,
\eeq
satisfy the symmetries of the Weyl and Riemann tensor, respectively.\footnote{Indeed, $\ti{\cC}_{abcd} \equiv \ti{\cC}_{cdab}$ is shown by contracting both sides with $\vep^{cdef}$ and $\ti{\cC}_{a[bcd]} \equiv 0$ is shown by contracting $\ti{\cC}_{abcd}$ with $\vep^{bcde}$. Note, however, that $\ti{\cC}_{abcd}$ is not traceless, although it is effectively made so in the action by being contracted with a traceless tensor.} As one could expect, a finite Immirzi parameter leads to a parity violating term $\sim \cC_{abcd}\, \ti{\cC}^{abcd}$, while in the parity-symmetric case $|\ga| \to \infty$ we recover the result of \cite{Zinoviev,BBB2015}
\beq
S^{\rm non-chiral}_{\rm (A)dS} = -\frac{3}{64 \pi G \La} \int \bar{e}\, \cC_{abcd} \, \cC^{cdab} + \Ord(\de A^3) \, .
\eeq
On the other hand, given the symmetries of $\ti{\cC}_{abcd}$ and the tracelessness of $\cC_{abcd}$, in the (anti-)self-dual cases $\ga = \pm i$ the action depends exclusively on the irreducible component $\what{\bm{\cS}}_+$
\beq 
S^{\rm (A)SD}_{\rm (A)dS} = -\frac{3}{64 \pi G \La} \int \bar{e}\, \cW^{\pm}_{abcd}\, \cW_{\pm}^{abcd} + \Ord(\de A^3) \, ,
\eeq
and only through the combination 
\beq
\cW^{\pm}_{abcd} := \frac{1}{\sqrt{2}} \[ \what{\cS}^+_{abcd} \pm \frac{i}{2}\, \vep_{ab}^{\,\,\,\,\,\,ef} \what{\cS}^+_{efcd} \]  \, .
\eeq
Finally, the fact that the action depends solely on the traceless component $\cC_{abcd}$, and not on $\cF_{ab}$, is a peculiarity of GR. To see this, let us consider the generalization of \eqref{eq:LPC} to an arbitrary Lagrangian function 
\beq \label{eq:Lf}
L = f(\bm{\cX}, \star) \, , 
\eeq 
that is a homogeneous function of $\bm{\cX}$ of degree one
\beq \label{eq:hom1}
f(c \bm{\cX},\star) \equiv c f(\bm{\cX},\star) \, .
\eeq
This requirement is essentially the condition of $L$ being a 4-form, or equivalently of having a Lagrangian density of weight one, i.e. it is needed for invariance under the full diffeomorphism group. Repeating the above procedure for \eqref{eq:Lf}, we first note that we can again focus on squares of $\bm{\cX}^{(1)}$, because the $\sim \bm{\cX}^{(2)}$ contributions correspond to topological terms. Then \eqref{eq:cXpert} implies that the result can only depend on the $\cS$, $\ti{\cS}$, $\what{\bm{\cS}}_+$ and $\bm{\cA}_-$ components of the perturbation $\de \bm{\cF}$. As we just saw, all of them depend exclusively on the fully traceless component $\cC_{abcd}$, except for $\cS \equiv \bra \de \bm{\cF} \ket \equiv \de \cF/2$. Thus, it is the fact that the $c$ function is independent of $\cS$ that makes the quadratic GR action special.

\section{No go for real modified actions}  \label{sec:ghost}

A remarkable fact about the SD formulation is that it admits an infinite parameter family of modifications with the same degrees of freedom as GR \cite{Bengtsson1990, Krasnov2006,Bengtsson2007,Krasnov2007,Krasnov2008b,Krasnov2008,Freidel2008,Krasnov2009,Krasnov2012}. In its pure connection formulation this is simply the SD analogue of \eqref{eq:Lf}, that is, all the Lagrangians of the form
\beq \label{eq:Lfc}
L = f(\bm{X}) \, , \hspace{1cm} X^{ij} := F^i \we F^j \, ,
\eeq
where $f$ is homogeneous of degree one, thus generalizing \eqref{eq:LK}. As for the reality constraints, a natural generalization of \eqref{eq:realconstrF} would be \cite{HKS2016}
\beq \label{eq:realconstrFc}
F^i \we \bar{F}^j = 0 \, , \hspace{1cm} {\rm Im}\, f(\bm{X}) = 0 \, ,
\eeq
where the latter makes again the action real. Unlike in the case of GR, however, these reality constraints \eqref{eq:realconstrFc} are not guaranteed to be compatible with the dynamics of \eqref{eq:Lfc} \cite{HKS2016} and, in fact, it has been claimed that they are not \cite{Krasnov2015}. If this is the case, then the analogous metric to \eqref{eq:UrbGR}
\beq \label{eq:UrbMod}
g_{\al\be} \sim \vep_{ijk} \, \vep^{\mu\nu\ro\si} F^i_{\al\mu} F^j_{\nu\ro} F^k_{\si\be} \, , \hspace{1cm} \sqrt{|g|}\, \ed^4 x = \frac{1}{m^4} \, f(\bm{X}) \, ,
\eeq
where $m$ is some mass scale, cannot be real Lorentzian throughout the evolution. There might exist different reality constraints than \eqref{eq:realconstrFc} that are compatible, and a different metric than \eqref{eq:UrbMod} that these constraints would make real Lorentzian, but if not, then these modified theories admit no real Lorentzian metric formulation and therefore would not count as theories of gravity. Nevertheless, prior to imposing the reality constraints, the Lagrangian \eqref{eq:Lfc} has the same degrees of freedom as \eqref{eq:LK}. Consequently, in the Euclidean case where $A^i$ is real, \eqref{eq:UrbMod} is real Euclidean and no constraints are required, these theories are genuine modified (Euclidean) gravity theories. 

Putting aside this yet undetermined aspect for the Lorentzian case, the existence of this family of modified massless spin-2 self-interactions is possible because the standard uniqueness theorem of GR assumes symmetry under parity transformations, which is not the case of theories based on the self-dual component alone \cite{Krasnov2014}. In contrast, therefore, the analogous modification in the real non-chiral case $L = f(\star \bm{\cX})$ has eight degrees of freedom in general \cite{AK2008}, namely, one massless graviton (2), one massive graviton (5) and a scalar (1) \cite{Speziale2010,BPS2012}, meaning this is actually a ``bigravity" theory. This degree of freedom count remains unchanged if we allow for real parity-violating terms, i.e. the general Lagrangian \eqref{eq:Lf}, because in the absence of imaginary factors both the self-dual and the anti-self-dual components will be present and thus generically active. Note that this is not true in the Euclidean case, because the corresponding irreducible representations of SO(4) are real, so one can project out one of the two gravitons using only real parameters \cite{BPS2012}. Note also that, at the level of the BF formulation, the corresponding generalization of \eqref{eq:LBF} takes the form
\beq \label{eq:LfB}
L = \bra B | F \ket - \frac{1}{2}\, V \( \bm{\cY}, \star \) \, ,
\eeq
for some ``potential" $V$ that is also homogeneous of degree one. Integrating out $| B \ket$ then leads to a Lagrangian of the form \eqref{eq:Lf}. 

Here we would like to point out that, in the real Lorentzian formulation of these modified theories, one of the two gravitons is necessarily a ``ghost", i.e. it has negative-definite kinetic energy. The corresponding QFT is therefore either unstable or non-unitary and, more generally, such a pathology casts doubt on whether the fully non-linear theory can lead to sensible dynamics, both at the classical and quantum levels. Note that in \cite{BPS2012} the focus was on the scalar, which is notoriously ghost-like in the case of standard bigravity \cite{BD1972}, and on ways to eliminate it from the spectrum, in analogy with standard bigravity where this is possible \cite{dRGT2011,HR2011a,HR2011b,HR2011c,HR2012}. However, if one of the two gravitons is a ghost too, then the only viable theory is the one with no scalar and no second graviton, that is, GR. Our argumentation will also show why the second graviton was not identified as a ghost in \cite{Speziale2010,BPS2012}.  

In order to prove our assertion, we first need to lay down some quick facts. Let us consider the linearized theory for the fluctuations $\de A^{ab}$ around some background solution $\bar{A}^{ab}$ with associated torsionless vierbein $\bar{e}^a$. Focusing on spin-2 excitations (gravitons), we note that these lie in symmetric 3-tensors and that one can form exactly two such fields out of $\de A^{ab}$
\beq \label{eq:hijhtij}
h^{ij} := \bar{e}^{\mu(i} \de A^{j)0}_{\mu} \, , \hspace{1cm} \ti{h}^{ij} := \frac{1}{2}\,\bar{e}^{\mu(i} \vep^{j)kl} \de A^{kl}_{\mu} \, .
\eeq
The equations of motion of the pure connection theory being second-order in derivatives, there are at most two gravitons. In the case of GR, the simplicity constraints imposed by $\bm{\psi}$ on $| B \ket$, and therefore ultimately on the dynamics of the connection, neutralize a combination of $h^{ij}$ and $\ti{h}^{ij}$, thus leaving us with a single graviton. From the viewpoint of the canonical formulation, this corresponds to the presence of second-class constraints \cite{AK2008,Alexandrov2000} which are therefore due to the particular structure of the action, not to its gauge symmetries. The necessity of such constraints can be understood from the fact that the gauge parameters of the theory do not contain symmetric 3-tensors, so the corresponding first-class constraints and possible gauge choices cannot completely eliminate components of the form \eqref{eq:hijhtij}. As already noted in the previous section, at the level of the linearized action \eqref{eq:SPClin}, the special structure of GR manifests itself in the fact that the dynamics only depend on part of the curvature components $\cF_{abcd}$, namely, the fully traceless tensor $\cC_{abcd}$.

Thus, in the non-GR cases where the Lagrangian does not have the specific structure \eqref{eq:LPC} or \eqref{eq:LBF}, one should generically expect both $h^{ij}$ and $\ti{h}^{ij}$ to contain degrees of freedom, as shown at the fully non-perturbative level in \cite{AK2008}. In contrast, these extra degrees of freedom cannot occur in the SD case, simply because one can only form a single symmetric 3-tensor out of the self-dual component of the connection alone $\bar{e}^{\mu(i} \de A^{j)}_{\mu}$. From the canonical viewpoint, SO$(3,\Cs)$ GR only contains first-class constraints that are related to the gauge symmetries. Since the latter are the same for all Lagrangians \eqref{eq:Lfc} and the number of reality constraints \eqref{eq:realconstrFc} is also the same, the degree of freedom count is the same as in GR.

Now note that $h^{ij}$ and $\ti{h}^{ij}$ are even and odd under parity, respectively. Thus, if the action is parity-symmetric, then its free part will be diagonal in $h^{ij}$ and $\ti{h}^{ij}$, while if it is parity-violating, the two gravitons will lie in two independent linear combinations of these two fields. In any case, however, the fact that $h^{ij}$ and $\ti{h}^{ij}$ correspond to a boost and a rotation component of $\de A^{ab}$, respectively, implies that the kinetic terms of the two independent combinations, along with their associated momenta, will enter with opposite signs in the free Hamiltonian. This is because the Killing form of the Lorentz algebra has split signature ${\rm diag}(-1,-1,-1,1,1,1)$ so, whatever the independent combinations of $h^{ij}$ and $\ti{h}^{ij}$ are, they cannot enter through a positive definite scalar product. As a result, even after taking into account all available constraints and all possible gauge fixings, the free Hamiltonian of the modified theories cannot be positive-definite. In particular, if one graviton has positive-definite kinetic energy, then the other one will have negative-definite kinetic energy. This does not occur in standard bigravity, because in that case one has full control over the kinetic terms of the two metric/vierbein fields. In contrast, here both spin-2 excitations are forced to enter through a single spin connection field, which incidentally forces them to have the opposite behavior under parity transformations, contrary to the two gravitons of standard bigravity. The reason why this important point was missed in the literature is probably the fact that the explicit studies of the linearized theories performed so far \cite{Speziale2010,BPS2012} were always carried out in Euclidean signature, where the Killing form has definite sign $\sim {\rm diag}(1,1,1,1,1,1)$, and that the Lorentzian case is usually approached through complex GR. 

The fact that the modifications \eqref{eq:Lf} are pathological in the real case is important for two reasons. First, it highlights in yet another way the special place GR holds in describing interacting spin-2 dynamics. Second, such modified theories have been considered as a starting point for unified gauge theories where the gauge group is extended to include other forces on top of gravity \cite{Smolin2007,LSS2010}. Since the purely gravitational theory is non-viable to begin with, such generalizations are bound to fail too, i.e. they lack the action structure that is required to eliminate the ghost graviton. Consequently, if one wishes to extend the group using a modified gravity theory, then only the SD theories \eqref{eq:Lfc} would seem appropriate, as considered in \cite{TGK2009,TGKS2010} for instance. In that case, however, there remains the difficulty of finding reality constraints that are compatible and that allow the construction of some real Lorentzian metric.

\section{Group extensions and unification} \label{eq:unif}

The fact that only a particular choice of Lagrangian function leads to sensible dynamics at the purely gravitational level constraints severely the set of potentially viable actions for the corresponding unified theories, i.e. those obtained through extension of the gauge group ${\rm SO}(1,3) \to G$. Indeed, if the $G = {\rm SO}(1,3)$ case has ghosts, then so will the $G \supset {\rm SO}(1,3)$ one. Conversely, extending the group with a Lagrangian that has the GR form has a chance of yielding a viable theory. So let us discuss some aspects of these theories.

We first observe that the presence of the $\star$ operator precludes us from extending the gauge group in general. The only exceptions are groups that also have such a complex structure, e.g. groups that are the complexification of some other group. This is also clear from the SL$(2,\Cs)$ form of the Lagrangian, i.e. \eqref{eq:LPC} with \eqref{eq:vepXC}, which is straightforwardly generalizable for groups with complex curvature coefficients. Note, however, that for the specific chiral choices $\ga = \pm 1$ we have $\bm{z}^2 = \pm \bm{1}$, so the pure connection Lagrangian \eqref{eq:LPC2} is independent of $\star$, and so is the BF one \eqref{eq:LBF2} if we restrict to $\bm{w} = w \bm{1}$. Observe also that the difference between the two $\ga = \pm 1$ choices amounts to flipping the sign of $\la$, so from now on we consider the case $\ga = 1$ for definiteness. Moreover, the right and wrong sectors too are now related by a simple sign flip of $\la$ (see \eqref{eq:GLaresc}).

With $|\ga| = 1$ the GR actions can be generalized straightforwardly to the ones of arbitrary group $G$ by simply keeping the same matrix/vector notation, only now the implicit indices are $A,B,C,\dots$ of the adjoint representation of $G$, with structure constants $f^A_{\,\,\,\,BC}$, and are displaced using the Killing form
\beq
\ka_{AB} \sim f^C_{\,\,\,\,DA} f^D_{\,\,\,\,CB} \, .
\eeq
Interestingly, with the choice $G = {\rm SO}(3)$, or equivalently SU(2), we obtain the pure-connection Lagrangian for Euclidean four-dimensional GR derived by Krasnov \cite{Krasnov2011a,Krasnov2011b} as a stepping stone for the SD Lorentzian one. We now consider the relevant case for unification $G \supset {\rm SO}(1,3)$, meaning in particular that $|G| := {\rm dim}(G) > 6$, and work with the $\ph,\bm{\psi}$-dependent BF formulation \eqref{eq:LBFpsi}, now reading
\beq \label{eq:LBFuni}
L = \bra B | F \ket - \frac{1}{2}\, \bra B | \bm{\psi} | B \ket + \ph \[ \Bra \frac{w^2 \bm{\psi}}{w^2 \bm{1} + \bm{\psi}} \Ket - \la \] \, .
\eeq
The equations of motion of $| B \ket$ and $\bm{\psi}$ read
\beq 
| F \ket = \bm{\psi} | B \ket \, , \hspace{1cm} \( w^2 \bm{1} + \bm{\psi} \) \bm{\cY} \( w^2 \bm{1} + \bm{\psi} \) = 2 w^4 \ph \bm{1} \, ,
\eeq
respectively, and with the former we can rewrite the latter as 
\beq \label{eq:KKeq}
| K \ket \bra K | = 2 w^4 \ph \bm{1} \, , \hspace{1cm} | K \ket := w^2 | B \ket + | F \ket \, .
\eeq
Invoking some arbitrary non-singular vierbein field $e^a$, defining
\beq
\ti{K}^A_{ab} := \frac{1}{2}\, (\sqrt{\star})_{ab}^{\,\,\,\,\,\,cd} e_c^{\mu} e_d^{\nu} K^A_{\mu\nu} \, , 
\eeq
and using a matrix/vector notation only for the $[ab]$ indices of that quantity, we can express \eqref{eq:KKeq} as
\beq \label{eq:KKeq2}
\bra \ti{K}^A | \ti{K}^B \ket\, e = - 2w^4 \ph\, \ka^{AB} \, .
\eeq
Now if we assume that $G$ is semi-simple, then there exists a generator basis such that $\ka^{AB}$ is diagonal with $\pm 1$ entries. In that case \eqref{eq:KKeq2} admits no solutions, because it states that there exist $|G| > 6$ six-dimensional vectors $| \ti{K}^A \ket \neq 0$ that are linearly independent with respect to the scalar product $\ka_{[ab][cd]}$. The gauge group of the Standard Model is semi-simple, so if it is to be unified with SO(1,3) within some larger group, one would expect the Jordan form of the corresponding $\ka^{AB}$ to contain at least one diagonal block with dimension larger than six, thus making again \eqref{eq:KKeq2} impossible to solve. At the level of the pure-connection \eqref{eq:LPC} and constraint-free BF \eqref{eq:LBF} formulations, this problem manifests itself as the impossibility of $\bm{\cX}$ and $\bm{\cY}$ to be invertible matrices for $|G| > 6$. They both take the respective forms $\bra F^A | F^B \ket$ and $\bra B^A | B^B \ket$, but the vectors forming the columns of these matrices cannot be linearly independent for $|G| > 6$. Note that the absence of classical solutions does not necessarily mean that the quantum theory does not exist, it only means that the path integral has no saddle point. Without a notion of ``on-shell", however, most of the perturbative mathematical methods fail, thus making the manipulation of such theories considerably more involved. 

This situation can be avoided by introducing an extra set of dynamical 0-forms $h^{\al}$, forming some representation $R$ of $G$, along with their corresponding covariant momenta, i.e. a set of 3-forms $\pi_{\al}$, given that we are working with first-order Lagrangians. Using the generators in this representation $(T_R^A)^{\al}_{\,\,\,\be}$, one could then form a symmetric matrix of 0-forms $\bm{z} \equiv \bm{z}(h)$ and consider for example Lagrangians that follow the GR structure \eqref{eq:LBFpsi}
\beq \label{eq:LBFuni2}
L = \bra B | F \ket + \pi_{\al} \we \ced h^{\al} - \frac{1}{2}\, \bra B | \bm{\psi} | B \ket + \ph \[ \Bra \bm{z}^2(h)\, \frac{w^2 \bm{\psi}}{w^2 \bm{1} + \bm{\psi}} \Ket - \la(h) \] \, .
\eeq
The analogous equation to \eqref{eq:KKeq} would then be
\beq
| K \ket \bra K | = 2 w^4 \ph\, \bm{z}^2(h) \, ,
\eeq
and would therefore admit solutions where $\bm{z}^2$ has rank six at most. In particular, the vacuum solution of physical interest, where space-time exists and the extra forces are trivial, would correspond to the VEV of $\bm{z}^2$ being some pure-Lorentz invariant matrix, thus providing the desired dynamical symmetry breaking mechanism $G \to {\rm SO}(1,3) \times G'$. Observe also that it is necessary to make $h^{\al}$ dynamical, because otherwise it would alter the $\bm{\psi}$-dependence of the action, thus introducing ghosts. In fact, although we have maintained the GR structure in this extension, it is not at all guaranteed that a Lagrangian of the form \eqref{eq:LBFuni2} maintains the desired second-class constraints, or that it is devoid of pathologies in general, and a dedicated study should be carried out. We leave this to future work.

Finally, note that the presence of fields like $h^{\al}$ and $\pi_{\al}$ is also welcome for another reason, namely, because they allow us to include the type of fields one encounters in the matter sector of the Standard Model. Indeed, on the one hand the Higgs field is a set of 0-forms, while 3-forms with non-zero VEV are needed in order to form first-order Lagrangians out of spinor 0-forms $\sim \pi \we \psi \ced \psi$.

\acknowledgments

The author is grateful to the referees for useful comments and suggestions. This work is supported by a Consolidator Grant of the European Research Council (ERC-2015-CoG grant 680886).


\begin{thebibliography}{20}   



\bibitem{KP2017}
  K.~Krasnov and R.~Percacci,   
  Class.\ Quant.\ Grav.\  {\bf 35} (2018) no.14,  143001, \href{http://xxx.lanl.gov/abs/1712.03061}{{\tt 1712.03061}}.


\bibitem{Eddington}
A.~S.~Eddington, The Mathematical Theory of Relativity (Cambridge University Press, 1920), Chap. 7.


\bibitem{Schrodinger}
E.~Schr\" odinger, Space-Time Structure (Cambridge University Press, 1950), Chap. 12.



\bibitem{Plebanski}
  J.~F.~Plebanski,  
  J.\ Math.\ Phys.\  {\bf 18} (1977) 2511.


\bibitem{CJD1989} 
  R.~Capovilla, T.~Jacobson and J.~Dell,   
  Phys.\ Rev.\ Lett.\  {\bf 63} (1989) 2325.


\bibitem{CJD1991}
  R.~Capovilla, T.~Jacobson and J.~Dell,   
  Class.\ Quant.\ Grav.\  {\bf 8} (1991) 59.


\bibitem{CJD1992}
  R.~Capovilla, T.~Jacobson and J.~Dell,   
  Class.\ Quant.\ Grav.\  {\bf 9} (1992) 1839.


\bibitem{Peldan}
  P.~Peldan,   
  Class.\ Quant.\ Grav.\  {\bf 11} (1994) 1087, \href{http://xxx.lanl.gov/abs/gr-qc/9305011}{{\tt gr-qc/9305011}}.


\bibitem{CJ1992}
  R.~Capovilla and T.~Jacobson,   
  Mod.\ Phys.\ Lett.\ A {\bf 7} (1992) 1871, \href{http://xxx.lanl.gov/abs/gr-qc/9207002}{{\tt gr-qc/9207002}}.



\bibitem{Krasnov2011a}
  K.~Krasnov,   
  Phys.\ Rev.\ D {\bf 84} (2011) 024034, \href{http://xxx.lanl.gov/abs/1101.4788}{{\tt 1101.4788}}.


\bibitem{Krasnov2011b}
  K.~Krasnov,  
  Phys.\ Rev.\ Lett.\  {\bf 106} (2011) 251103, \href{http://xxx.lanl.gov/abs/1103.4498}{{\tt 1103.4498}}.



\bibitem{DKS2012}
  G.~Delfino, K.~Krasnov and C.~Scarinci,   
  JHEP {\bf 1503} (2015) 118, \href{http://xxx.lanl.gov/abs/1205.7045}{{\tt 1205.7045}}.
 

\bibitem{DKS2012b}
  G.~Delfino, K.~Krasnov and C.~Scarinci,   
  JHEP {\bf 1503} (2015) 119, \href{http://xxx.lanl.gov/abs/1210.6215}{{\tt 1210.6215}}.
 

\bibitem{GKS2013}
  K.~Groh, K.~Krasnov and C.~F.~Steinwachs,   
  JHEP {\bf 1307} (2013) 187, \href{http://xxx.lanl.gov/abs/1304.6946}{{\tt 1304.6946}}.


\bibitem{DKS2014}
  G.~Delfino, K.~Krasnov and C.~Scarinci,   
  JHEP {\bf 1503} (2015) 120, \href{http://xxx.lanl.gov/abs/1410.5647}{{\tt 1410.5647}}.




\bibitem{DPF1998}
  R.~De Pietri and L.~Freidel,   
  Class.\ Quant.\ Grav.\  {\bf 16} (1999) 2187, \href{http://xxx.lanl.gov/abs/gr-qc/9804071}{{\tt gr-qc/9804071}}.


\bibitem{HK2015}
  Y.~Herfray and K.~Krasnov,   
\href{http://xxx.lanl.gov/abs/1503.08640}{{\tt 1503.08640}}.  



\bibitem{Krasnov2017}
  K.~Krasnov,   
  Class.\ Quant.\ Grav.\  {\bf 35} (2018) no.14,  147001, \href{http://xxx.lanl.gov/abs/1708.07694}{{\tt 1708.07694}}.





\bibitem{Zinoviev}
  Y.~M.~Zinoviev,   
  \href{http://xxx.lanl.gov/abs/hep-th/0504210}{{\tt hep-th/0504210}}.

\bibitem{BBB2015}
  T.~Basile, X.~Bekaert and N.~Boulanger,   
  Phys.\ Rev.\ D {\bf 93} (2016) no.12,  124047, \href{http://xxx.lanl.gov/abs/1512.09060}{{\tt 1512.09060}}.






\bibitem{Bengtsson1990}
  I.~Bengtsson,   
  Phys.\ Lett.\ B {\bf 254} (1991) 55.


\bibitem{Krasnov2006}
  K.~Krasnov,   
   \href{http://xxx.lanl.gov/abs/hep-th/0611182}{{\tt hep-th/0611182}}.  


\bibitem{Bengtsson2007}
  I.~Bengtsson,   
  Mod.\ Phys.\ Lett.\ A {\bf 22} (2007) 1643, \href{http://xxx.lanl.gov/abs/gr-qc/0703114}{{\tt gr-qc/0703114}}.  



\bibitem{Krasnov2007}
  K.~Krasnov,
  Mod.\ Phys.\ Lett.\ A {\bf 22} (2007) 3013, \href{http://xxx.lanl.gov/abs/0711.0697}{{\tt 0711.0697}}.  


\bibitem{Krasnov2008b}
  K.~Krasnov,   
  Phys.\ Rev.\ Lett.\  {\bf 100} (2008) 081102, \href{http://xxx.lanl.gov/abs/0711.0090}{{\tt 0711.0090}}.  


\bibitem{Krasnov2008}
  K.~Krasnov,
  Class.\ Quant.\ Grav.\  {\bf 26} (2009) 055002, \href{http://xxx.lanl.gov/abs/0811.3147}{{\tt 0811.3147}}.


\bibitem{Freidel2008}
  L.~Freidel,
\href{http://xxx.lanl.gov/abs/0812.3200}{{\tt 0812.3200}}.  



\bibitem{Krasnov2009}
  K.~Krasnov,
  EPL {\bf 89} (2010) no.3,  30002, \href{http://xxx.lanl.gov/abs/0910.4028}{{\tt 0910.4028}}.


\bibitem{Krasnov2012}
  K.~Krasnov,   
  Proc.\ Roy.\ Soc.\ Lond.\ A {\bf 468} (2012) 2129, \href{http://xxx.lanl.gov/abs/1202.6183}{{\tt 1202.6183}}.










\bibitem{CMPR2001}
  R.~Capovilla, M.~Montesinos, V.~A.~Prieto and E.~Rojas, 
  Class.\ Quant.\ Grav.\  {\bf 18} (2001) L49, \href{http://xxx.lanl.gov/abs/gr-qc/0102073}{{\tt gr-qc/0102073}}.


\bibitem{FS2012}
  L.~Freidel and S.~Speziale,   
  SIGMA {\bf 8} (2012) 032, \href{http://xxx.lanl.gov/abs/1201.4247}{{\tt 1201.4247}}.

 

\bibitem{SS2009}
  L.~Smolin and S.~Speziale, 
  Phys.\ Rev.\ D {\bf 81} (2010) 024032, \href{http://xxx.lanl.gov/abs/0908.3388}{{\tt 0908.3388}}.



\bibitem{MV2010}
  M.~Montesinos and M.~Velazquez,  
  Phys.\ Rev.\ D {\bf 81} (2010) 044033, \href{http://xxx.lanl.gov/abs/1002.3836}{{\tt 1002.3836}}.


\bibitem{Holst1995}
  S.~Holst,   
  Phys.\ Rev.\ D {\bf 53} (1996) 5966, \href{http://xxx.lanl.gov/abs/gr-qc/9511026}{{\tt gr-qc/9511026}}.


\bibitem{Alexandrov2000}
  S.~Alexandrov,   
  Class.\ Quant.\ Grav.\  {\bf 17} (2000) 4255, \href{http://xxx.lanl.gov/abs/gr-qc/0005085}{{\tt gr-qc/0005085}}.


\bibitem{Livine2006}
  E.~R.~Livine,  
  In *Oriti, D. (ed.): Approaches to quantum gravity* (Cambridge University Press, 2009), p. 253-271, \href{http://xxx.lanl.gov/abs/gr-qc/0608135}{{\tt gr-qc/0608135}}.



\bibitem{MRC2017}
  M.~Montesinos, J.~Romero and M.~Celada,   
  Phys.\ Rev.\ D {\bf 97} (2018) no.2,  024014, \href{http://xxx.lanl.gov/abs/1712.00040}{{\tt 1712.00040}}.


  
  
\bibitem{HKS2016}
  Y.~Herfray, K.~Krasnov and Y.~Shtanov,   
  Class.\ Quant.\ Grav.\  {\bf 33} (2016) 235001,  \href{http://xxx.lanl.gov/abs/1510.05820}{{\tt 1510.05820}}.



\bibitem{Urbantke}
  H.~Urbantke,   
  J.\ Math.\ Phys.\  {\bf 25} (1984) no.7,  2321.




\bibitem{Krasnov2015}
  K.~Krasnov,   
  Class.\ Quant.\ Grav.\  {\bf 33} (2016) no.15,  155012, \href{http://xxx.lanl.gov/abs/1512.07110}{{\tt 1512.07110}}.



\bibitem{Krasnov2014}
  K.~Krasnov,   
  JHEP {\bf 1510} (2015) 037, \href{http://xxx.lanl.gov/abs/1410.8006}{{\tt 1410.8006}}.




\bibitem{AK2008}
  S.~Alexandrov and K.~Krasnov,   
  Class.\ Quant.\ Grav.\  {\bf 26} (2009) 055005, \href{http://xxx.lanl.gov/abs/0809.4763}{{\tt 0809.4763}}.



\bibitem{Speziale2010}
  S.~Speziale,   
  Phys.\ Rev.\ D {\bf 82} (2010) 064003, \href{http://xxx.lanl.gov/abs/1003.4701}{{\tt 1003.4701}}.

 
 \bibitem{BPS2012}
  D.~Beke, G.~Palmisano and S.~Speziale,   
  JHEP {\bf 1203} (2012) 069, \href{http://xxx.lanl.gov/abs/1112.4051}{{\tt 1112.4051}}.


\bibitem{BD1972}
  D.~G.~Boulware and S.~Deser,   
  Phys.\ Rev.\ D {\bf 6} (1972) 3368.


\bibitem{dRGT2011}
  C.~de Rham, G.~Gabadadze and A.~J.~Tolley,   
  Phys.\ Rev.\ Lett.\  {\bf 106} (2011) 231101, \href{http://xxx.lanl.gov/abs/1011.1232}{{\tt 1011.1232}}.


\bibitem{HR2011a}
  S.~F.~Hassan and R.~A.~Rosen,   
  Phys.\ Rev.\ Lett.\  {\bf 108} (2012) 041101, \href{http://xxx.lanl.gov/abs/1106.3344}{{\tt 1106.3344}}.


\bibitem{HR2011b}
  S.~F.~Hassan and R.~A.~Rosen,   
  JHEP {\bf 1202} (2012) 126, \href{http://xxx.lanl.gov/abs/1109.3515}{{\tt 1109.3515}}.


\bibitem{HR2011c}
  S.~F.~Hassan and R.~A.~Rosen,  
  JHEP {\bf 1204} (2012) 123, \href{http://xxx.lanl.gov/abs/1111.2070}{{\tt 1111.2070}}.


\bibitem{HR2012}
  K.~Hinterbichler and R.~A.~Rosen,  
  JHEP {\bf 1207} (2012) 047, \href{http://xxx.lanl.gov/abs/1203.5783}{{\tt 1203.5783}}.




\bibitem{Smolin2007}
  L.~Smolin,   
  Phys.\ Rev.\ D {\bf 80} (2009) 124017, \href{http://xxx.lanl.gov/abs/0712.0977}{{\tt 0712.0977}}.

\bibitem{LSS2010} 
  A.~G.~Lisi, L.~Smolin and S.~Speziale,   
  J.\ Phys.\ A {\bf 43} (2010) 445401, \href{http://xxx.lanl.gov/abs/1004.4866}{{\tt 1004.4866}}.
 


\bibitem{TGK2009}
  A.~Torres-Gomez and K.~Krasnov,   
  Phys.\ Rev.\ D {\bf 81} (2010) 085003, \href{http://xxx.lanl.gov/abs/0911.3793}{{\tt 0911.3793}}.

\bibitem{TGKS2010}
  A.~Torres-Gomez, K.~Krasnov and C.~Scarinci,   
  Phys.\ Rev.\ D {\bf 83} (2011) 025023, \href{http://xxx.lanl.gov/abs/1011.3641}{{\tt 1011.3641}}.








\end{thebibliography}
\end{document}